\documentclass[conference,a4paper]{IEEEtran}
\usepackage[dvips]{graphicx}
\usepackage{setspace,psfrag,cite,subfigure, amsfonts,theorem}
\usepackage{amssymb}
\usepackage{bm}
\usepackage[french,english]{babel}
\usepackage{color}
\usepackage{amsmath}
\usepackage{algorithmic}
\usepackage{algorithm}

\ifCLASSINFOpdf
\else
\fi

\begin{document}
%

\title{On Sparsity Averaging}

\author{%
\IEEEauthorblockN{
Rafael E. Carrillo\IEEEauthorrefmark{1}, 
Jason D. McEwen\IEEEauthorrefmark{2}, 
and Yves Wiaux\IEEEauthorrefmark{1}\IEEEauthorrefmark{3}\IEEEauthorrefmark{4}}
\IEEEauthorblockA{\IEEEauthorrefmark{1} 
Institute of Electrical Engineering, Ecole Polytechnique F{\'e}d{\'e}rale de Lausanne (EPFL),
      CH-1015 Lausanne, Switzerland.}
\IEEEauthorblockA{\IEEEauthorrefmark{2}
Department of Physics and Astronomy, University College London, London WC1E 6BT, UK.}
\IEEEauthorblockA{\IEEEauthorrefmark{3}
Department of Radiology and Medical Informatics, University of Geneva (UniGE), 
      CH-1211 Geneva, Switzerland.}
\IEEEauthorblockA{\IEEEauthorrefmark{4}
Department of Radiology, Lausanne University Hospital (CHUV), CH-1011 Lausanne, Switzerland.
}
}

\maketitle

\begin{abstract}
Recent developments in \cite{carrillo12} and \cite{carrillo13} introduced a novel regularization method for compressive imaging in the context of compressed sensing with coherent redundant dictionaries. The approach relies on the observation that natural images exhibit strong \emph{average sparsity} over multiple coherent frames. The associated reconstruction algorithm, based on an \emph{analysis} prior and a \emph{reweighted} $\ell_1$ scheme, is dubbed Sparsity Averaging Reweighted Analysis (SARA). We review these advances and extend associated simulations establishing the superiority of SARA to regularization methods based on sparsity in a single frame, for a generic spread spectrum acquisition and for a Fourier acquisition of particular interest in radio astronomy.
\end{abstract}

\IEEEpeerreviewmaketitle

\section{Introduction}
\label{sec:intro}
Consider a complex-valued signal $\bm{x}\in\mathbb{C}^{N}$, assumed to be sparse in some orthonormal basis $\mathsf{\Psi}\in\mathbb{C}^{N\times N}$, and also consider the measurement model $\bm{y}=\mathsf{\Phi}\bm{x}+\bm{n}$, where $\bm{y}\in\mathbb{C}^{M}$ denotes the measurement vector, $\mathsf{\Phi}\in\mathbb{C}^{M\times N}$ with $M<N$ is the sensing matrix and $\bm{n}\in\mathbb{C}^{M}$ represents the observation noise. The most common approach in compressed sensing (CS) is to recover $\bm{x}$ from $\bm{y}$ solving the following convex problem \cite{fornasier11}:
\begin{equation}
\min_{\bar{\bm{\alpha}}\in\mathbb{C}^{N}}\|\bar{\bm{\alpha}}\|_{1}
\textnormal{ subject to }\| \bm{y}-\mathsf{\Phi \Psi}\bar{\bm{\alpha}}\|_{2}\leq\epsilon,\label{cs1}
\end{equation}
where $\epsilon$ is an upper bound on the $\ell_{2}$ norm of the noise and $\|\cdot\|_1$ denotes the $\ell_{1}$ norm of a complex-valued vector. The signal is recovered as $\hat{\bm{x}}=\mathsf{\Psi}\hat{\bm{\alpha}}$, where $\hat{\bm{\alpha}}$ denotes the solution to \eqref{cs1}. Such problems that solve for the representation of the signal in a sparsity basis are known as synthesis-based problems. The standard CS theory provides results for the recovery of $\bm{x}$ from $\bm{y}$ if $\mathsf{\Phi}$ obeys a Restricted Isometry Property (RIP) and $\mathsf{\Psi}$ is orthonormal~\cite{fornasier11}. However, signals often exhibit better sparsity in an overcomplete dictionary  \cite{gribonval03,bobin07,starck10}. 

Recent works have begun to address the case of CS with redundant dictionaries. In this setting the signal $\bm{x}$ is expressed in terms of a dictionary $\mathsf{\Psi}\in\mathbb{C}^{N\times D}$, $N<D$, as $\bm{x} = \mathsf{\Psi}\bm{\alpha}$, $\bm{\alpha}\in\mathbb{C}^{D}$. Rauhut et al.~\cite{rauhut08} find conditions on the dictionary $\mathsf{\Psi}$ such that the compound matrix $\mathsf{\Phi \Psi}$ obeys the RIP to accurately recover $\bm{\alpha}$ by solving a synthesis-based problem. Cand\`{e}s et al.~\cite{candes10} provide a theoretical analysis of the $\ell_1$ analysis-based problem. As opposed to synthesis-based problems, analysis-based problems recover the signal itself solving:
\begin{equation}\label{cs6}
\min_{\bar{\bm{x}}\in\mathbb{C}^{N}}\|\mathsf{\Psi}^{\dagger}\bar{\bm{x}}\|_{1}
\textnormal{ subject to }\| \bm{y}-\mathsf{\Phi}\bar{\bm{x}}\|_{2}\leq\epsilon,
\end{equation}
where $\mathsf{\Psi}^{\dagger}$ denotes the adjoint operator of $\mathsf{\Psi}$. The aforementioned work~\cite{candes10} extends the standard CS theory to coherent and redundant dictionaries, providing theoretical stability guarantees based on a general condition of the sensing matrix $\mathsf{\Phi}$, coined the Dictionary Restricted Isometry Property (D-RIP). 

In \cite{carrillo12} and \cite{carrillo13}, we proposed a novel sparsity analysis prior for compressive imaging in the context of CS with coherent and redundant dictionaries, relying on the observation that natural images are simultaneously sparse in various frames, in particular wavelet frames, or in their gradient. Promoting \emph{average sparsity} over multiple frames, as opposed to single frame sparsity, is an extremely powerful prior. The associated reconstruction algorithm, based on an \emph{analysis} prior and a \emph{reweighted} $\ell_1$ scheme, is dubbed Sparsity Averaging Reweighted Analysis (SARA)\footnote{In \cite{arberet13}, similar ideas were applied to the reverberant audio source separation problem exploiting sparsity in a redundant short time Fourier transform.}.

In this work, we review and further discuss these recent advances. The superiority of SARA to regularization methods based on sparsity in a single frame, as established through simulations for a generic spread spectrum acquisition, is described with an additional extensive visual support. Moreover, we bring a novel illustration for a realistic continuous Fourier sampling strategy of particular interest for radio interferometry in astronomy. We finally discuss possible avenues to establish explicit theoretical stability results for the algorithm.

\section{Sparsity Averaging Reweighted Analysis}
\label{sec:SARA}

Natural images are often complicated and encompass several types of structures admitting sparse representations in different frames. For example, piecewise smooth structures exhibit gradient sparsity, while extended structures are better encapsulated in wavelet frames. Observing that natural images actually exhibit sparsity in multiple frames, we hypothesise in \cite{carrillo12} and \cite{carrillo13} that average sparsity over multiple coherent frames represents a strong prior. We thus proposed the use of a dictionary composed of a concatenation of $q$ frames, i.e.
\begin{equation}\label{dict}
\mathsf{\Psi}=\frac{1}{\sqrt{q}}[\mathsf{\Psi}_1, \mathsf{\Psi}_2, \ldots, \mathsf{\Psi}_q],
\end{equation}
with $\mathsf{\Psi}\in\mathbb{C}^{N\times D}$, $N<D$, and an analysis $\ell_0$ prior, 
\begin{equation}\label{avs}
\|\mathsf{\Psi}^{\dagger}\bar{\bm{x}}\|_{0} \sim \frac{1}{q}\sum_{i=1}^q \|\mathsf{\Psi}_i^{\dagger}\bar{\bm{x}}\|_{0},
\end{equation}
to promote this average sparsity. Note that in this setting each frame contains all the signal information as opposed to component separation approaches such as \cite{gribonval03} and \cite{bobin07}. Also note on a theoretical level that a single signal cannot be arbitrarily sparse simultaneously in a set of incoherent frames. For example, a signal extremely sparse in the Dirac basis is completely spread in the Fourier basis. As discussed in \cite{carrillo13}, each frame, $\mathsf{\Psi}_i$, should be highly coherent with the other frames in order for the signal to have a sparse representation in $\mathsf{\Psi}$. Concatenation of the first eight orthonormal Daubechies wavelet bases (Db1-Db8) is an example of interest. The first Daubechies wavelet basis, Db1, is the Haar wavelet basis. It can be used as an alternative to gradient sparsity, usually imposed by a total variation (TV) prior, to promote piecewise smooth signals. The Db2-Db8 bases provide smoother decompositions. Coherence between the bases is ensured by the compact support of the Daubechies wavelets.

A reweighted $\ell_1$ minimization scheme \cite{candes08a} promotes average sparsity through the prior \eqref{avs}. The algorithm replaces the $\ell_0$ norm by a weighted $\ell_1$ norm and solves a sequence of weighted $\ell_1$ problems with weights essentially the inverse of the values of the solution of the previous problem: 
\begin{equation}\label{delta}
\min_{\bar{\bm{x}}\in\mathbb{C}^{N}}\|\mathsf{W\Psi}^{\dagger}\bar{\bm{x}}\|_{1}
\textnormal{ subject to }\| \bm{y}-\mathsf{\Phi}\bar{\bm{x}}\|_{2}\leq\epsilon,
\end{equation}
where $\mathsf{W}\in\mathbb{R}^{D\times D}$ is a diagonal matrix with positive weights. The solution to \eqref{delta} is denoted as $\Delta(\bm{y}, \mathsf{\Phi},\mathsf{W},\epsilon)$. We update the weights at each iteration, i.e.~after solving a complete weighted $\ell_1$ problem, by the function $f(\gamma,a) \propto(\gamma+|a|)^{-1}$, where $a$ denotes the coefficient value estimated at the previous iteration and $\gamma$ plays the role of a stabilization parameter, avoiding undefined weights when the signal value is zero. Note that as $\gamma\rightarrow 0$ the solution of the weighted $\ell_1$ problem approaches the solution of the $\ell_0$ problem. We use a homotopy strategy and solve a sequence of weighted $\ell_1$ problems with a decreasing sequence $\{\gamma^{(t)}\}$, with $t$ denoting the iteration time variable. 

The sparsity averaging reweighted analysis (SARA) algorithm is defined in Algorithm~\ref{alg1}, with $\mathsf{\Psi}$ defined as in \eqref{dict}. A rate parameter $\beta\in(0,1)$ controls the decrease of the sequence through $\gamma^{(t)}=\beta\gamma^{(t-1)}$. 
However, the noise standard deviation $\sigma_{\alpha}$ in the representation domain, rough estimate for a baseline above which significant signal components could be identified, serves as a lower bound: $\gamma^{(t)}\geq\sigma_{\alpha}=\sqrt{M/D}\sigma_n$, with $\sigma_n$ the noise standard deviation in measurement space. As a starting point we set $\hat{\bm{x}}^{(0)}$ as the solution of the $\ell_1$ problem and $\gamma^{(0)}=\sigma_s\left(\mathsf{\Psi}^{\dagger}\hat{\bm{x}}^{(0)}\right)$, where $\sigma_s(\cdot)$ takes the empirical standard deviation of a signal. The re-weighting process ideally stops  when the relative variation between successive solutions is smaller than some bound $\eta\in(0,1)$, or after the maximum number of iterations allowed, $N_{\rm{max}}$, is reached. We fix $\eta=10^{-3}$ and $\beta=10^{-1}$. 

\begin{algorithm}[h!]
\caption{SARA algorithm}\label{alg1}
\begin{algorithmic}[1]
\REQUIRE $\bm{y}$, $\mathsf{\Phi}$, $\epsilon$, $\sigma_{\alpha}$, $\beta$, $\eta$ and $N_{\rm{max}}$.
\ENSURE Reconstructed image $\hat{\bm{x}}$.
\STATE Initialize $t=1$, $\mathsf{W}^{(0)}=\mathsf{I}$ and $\rho=1$.
\STATE Compute\\
$\hat{\bm{x}}^{(0)}=\Delta(\bm{y}, \mathsf{\Phi},\mathsf{W}^{(0)},\epsilon)$,
$\gamma^{(0)}=\sigma_s\left(\mathsf{\Psi}^{\dagger}\hat{\bm{x}}^{(0)}\right)$.
\WHILE{$\rho>\eta$ and $t<N_{\rm{max}}$}
\STATE Update 
$\mathsf{W}_{ij}^{(t)}=f\left (\gamma^{(t-1)},\hat{\alpha}_{i}^{(t-1)}\right)\delta_{ij}$, \\
for $i,j=1,\ldots,D$ with $\hat{\bm{\alpha}}^{(t-1)}=\mathsf{\Psi}^{\dagger}\hat{\bm{x}}^{(t-1)}$.
\STATE Compute a solution
$\hat{\bm{x}}^{(t)}=\Delta(\bm{y}, \mathsf{\Phi},\mathsf{W}^{(t)},\epsilon)$.\\
\STATE Update $\gamma^{(t)}=\max\{\beta\gamma^{(t-1)},\sigma_{\alpha}\}$.
\STATE Update $\rho=\| \hat{\bm{x}}^{(t)}-\hat{\bm{x}}^{(t-1)}\|_2/\|\hat{\bm{x}}^{(t-1)}\|_2$.
\STATE $t\leftarrow t+1$
\ENDWHILE
\end{algorithmic}
\end{algorithm}

\section{Simulations}

\begin{figure}
\centering
   
    \includegraphics[trim = 3.9cm 1.1cm 3cm 1cm, clip, keepaspectratio, height = 3.8cm]{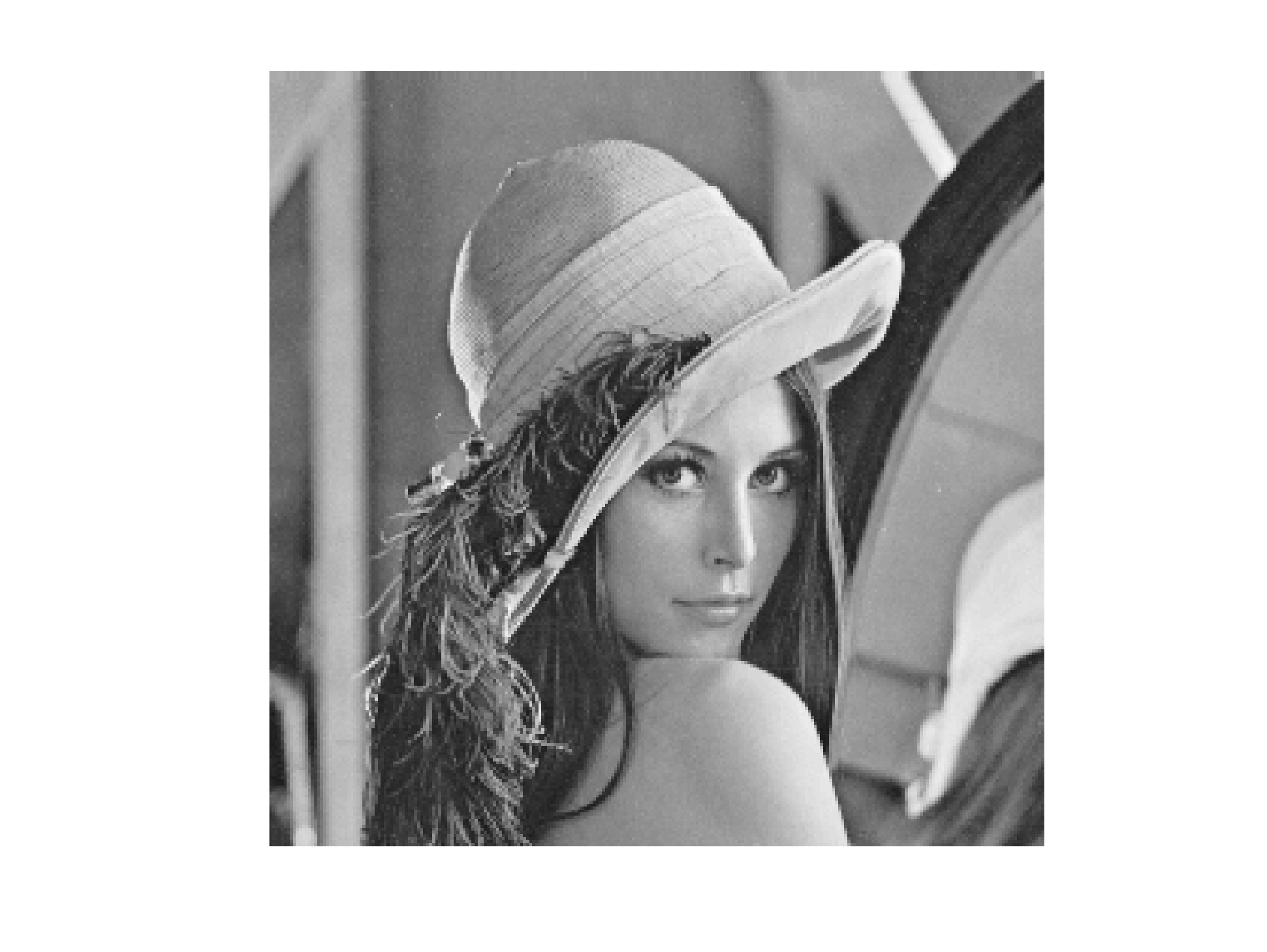}
    \includegraphics[trim = 0.8cm 0.1cm 1.5cm 0.8cm, clip, keepaspectratio, height = 3.8cm]{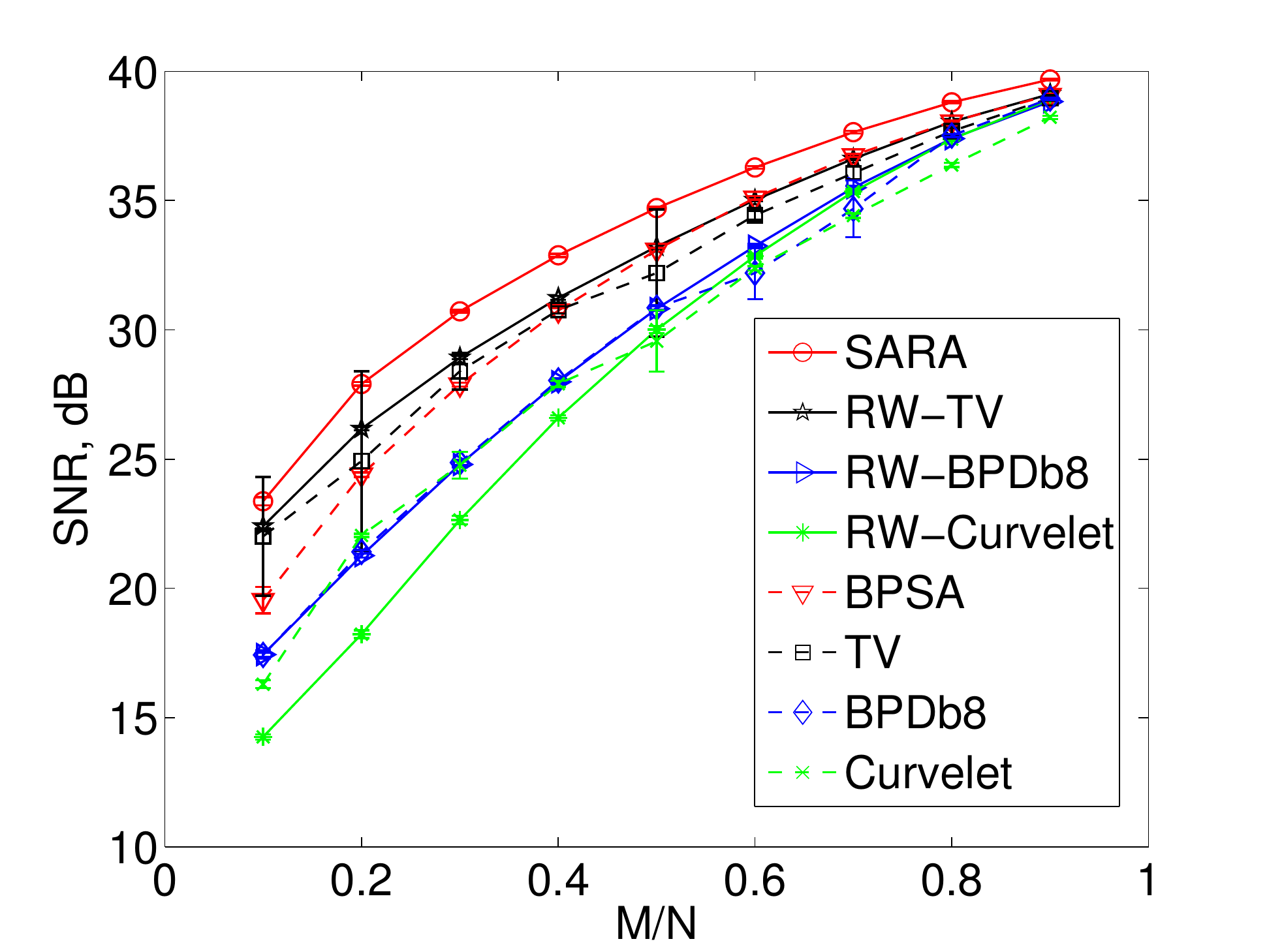}

\caption{Reconstruction quality results for Lena in the context of a spread spectrum acquisition. Left: original image. Right: SNR results against the undersampling ratio for an input SNR of 30 dB (average values over 100 simulations are shown with corresponding standard deviations).}
\label{fig:1}
\end{figure}

\begin{figure*}[t]

\centering
   
    \includegraphics[trim = 3.9cm 1.1cm 3cm 1cm, clip, keepaspectratio, height = 3.8cm]{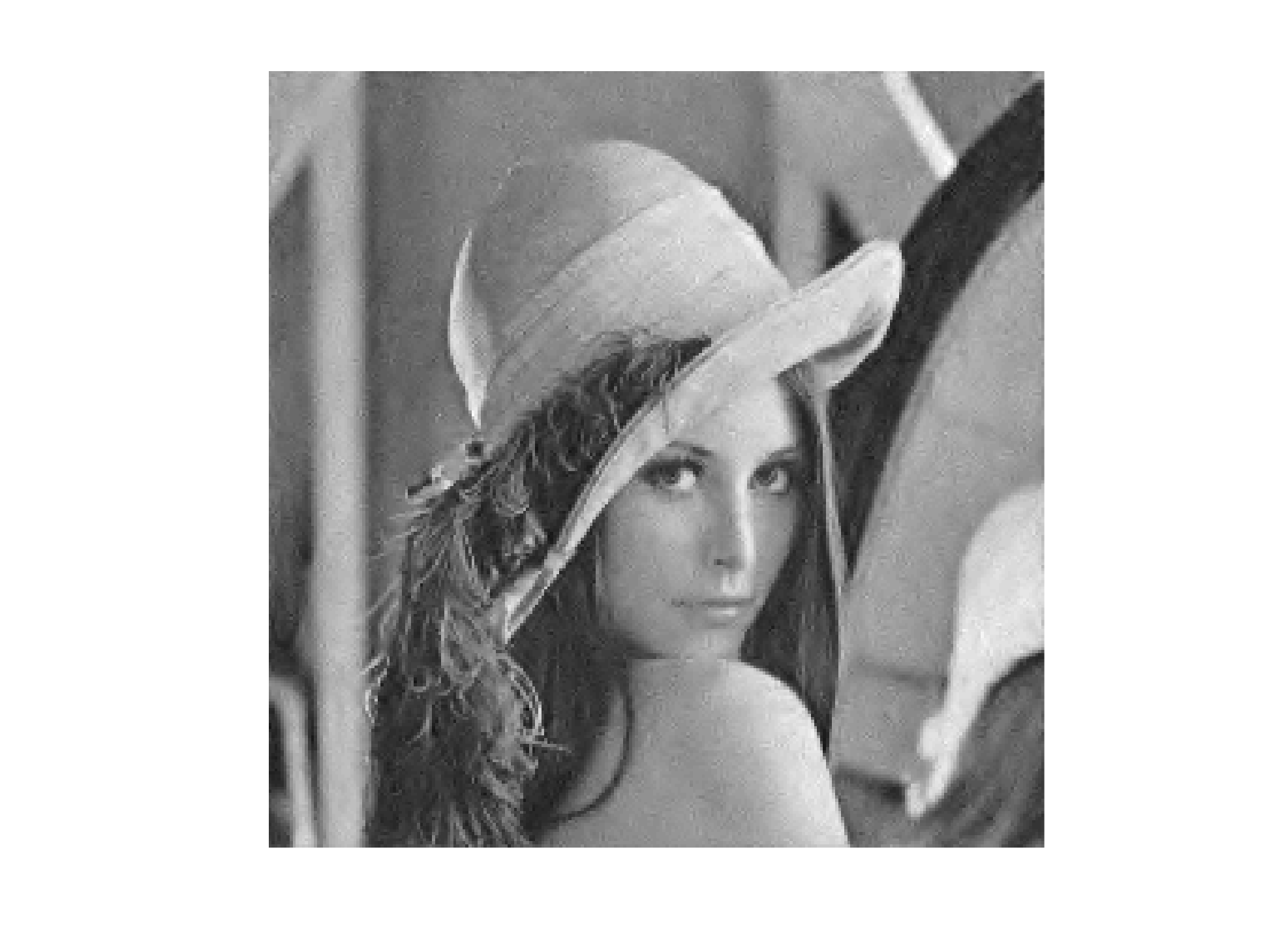}
    \includegraphics[trim = 3.9cm 1.1cm 2cm 1cm, clip, keepaspectratio, height = 3.8cm]{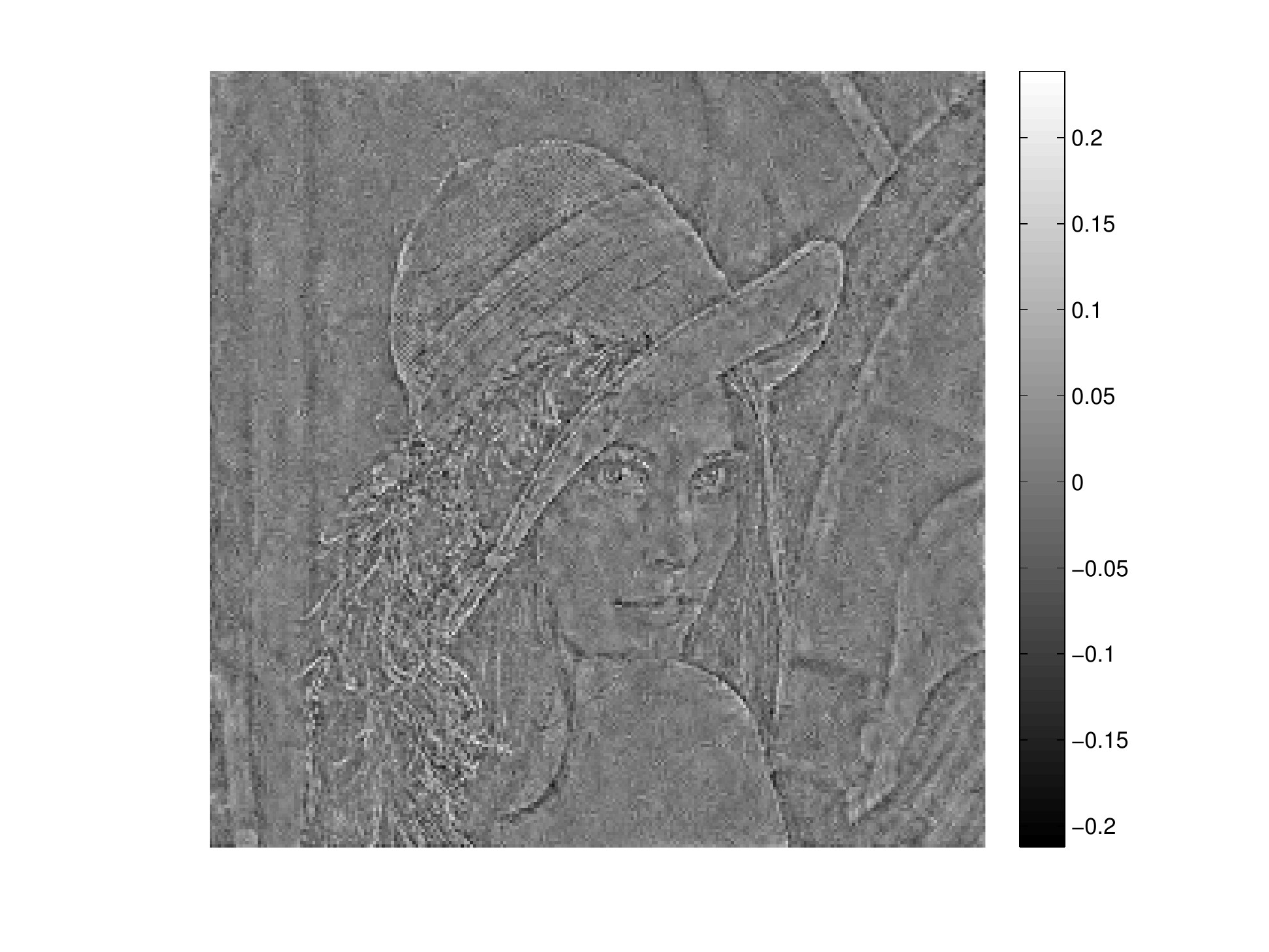}
    \includegraphics[trim = 3.9cm 1.1cm 3cm 1cm, clip, keepaspectratio, height = 3.8cm]{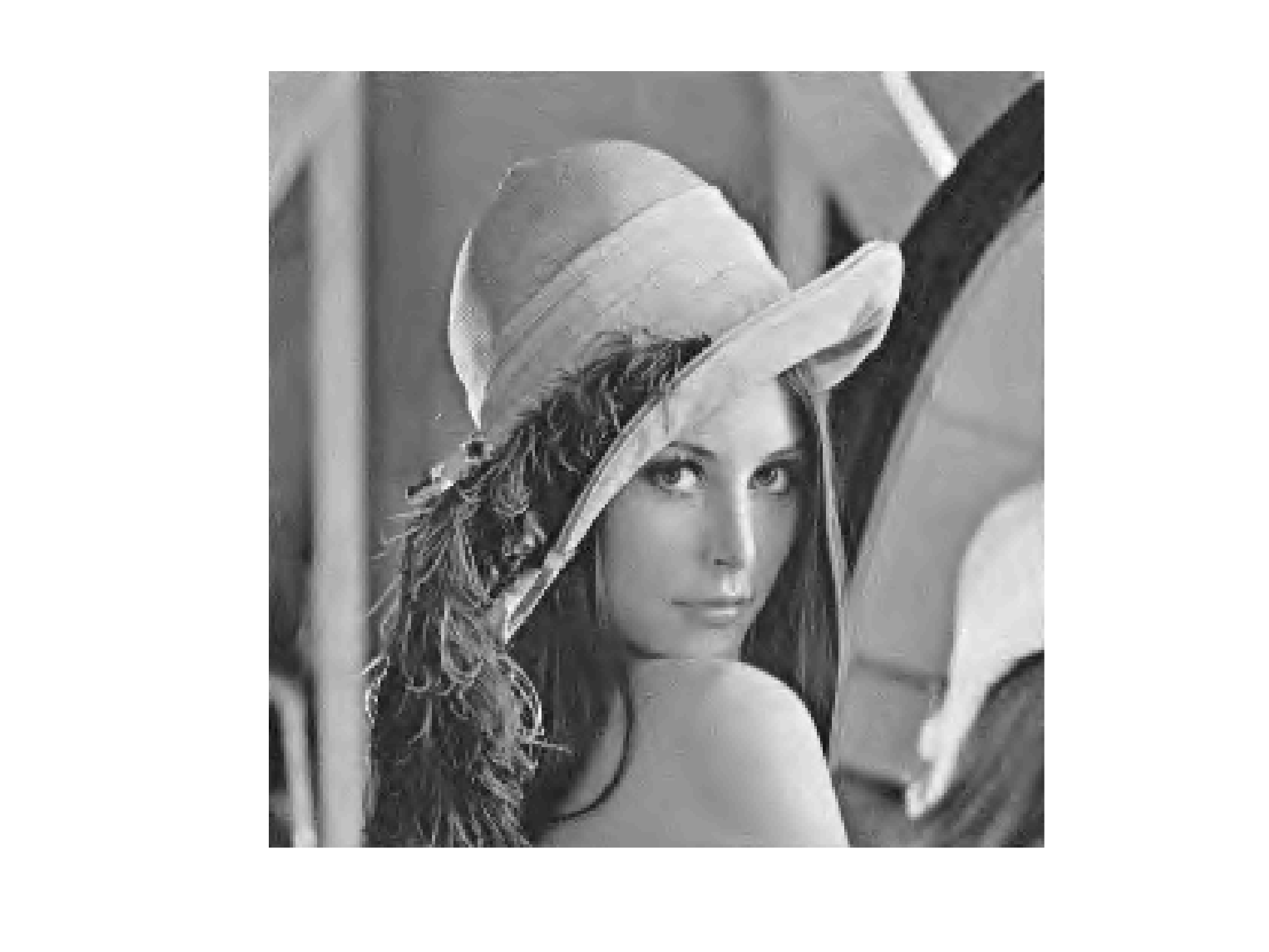}
    \includegraphics[trim = 3.9cm 1.1cm 2cm 1cm, clip, keepaspectratio, height = 3.8cm]{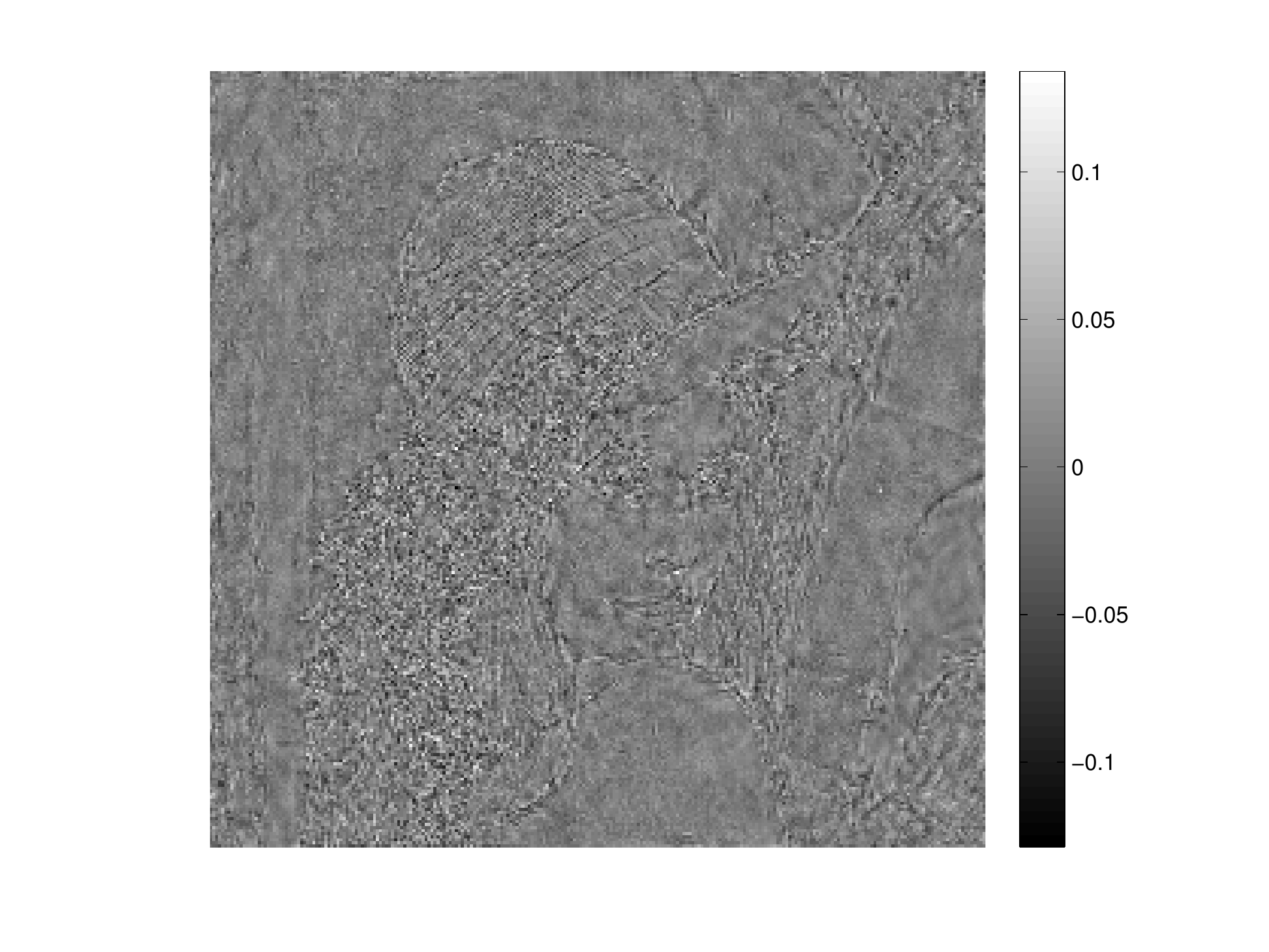}    
    
    \includegraphics[trim = 3.9cm 1.1cm 3cm 1cm, clip, keepaspectratio, height = 3.8cm]{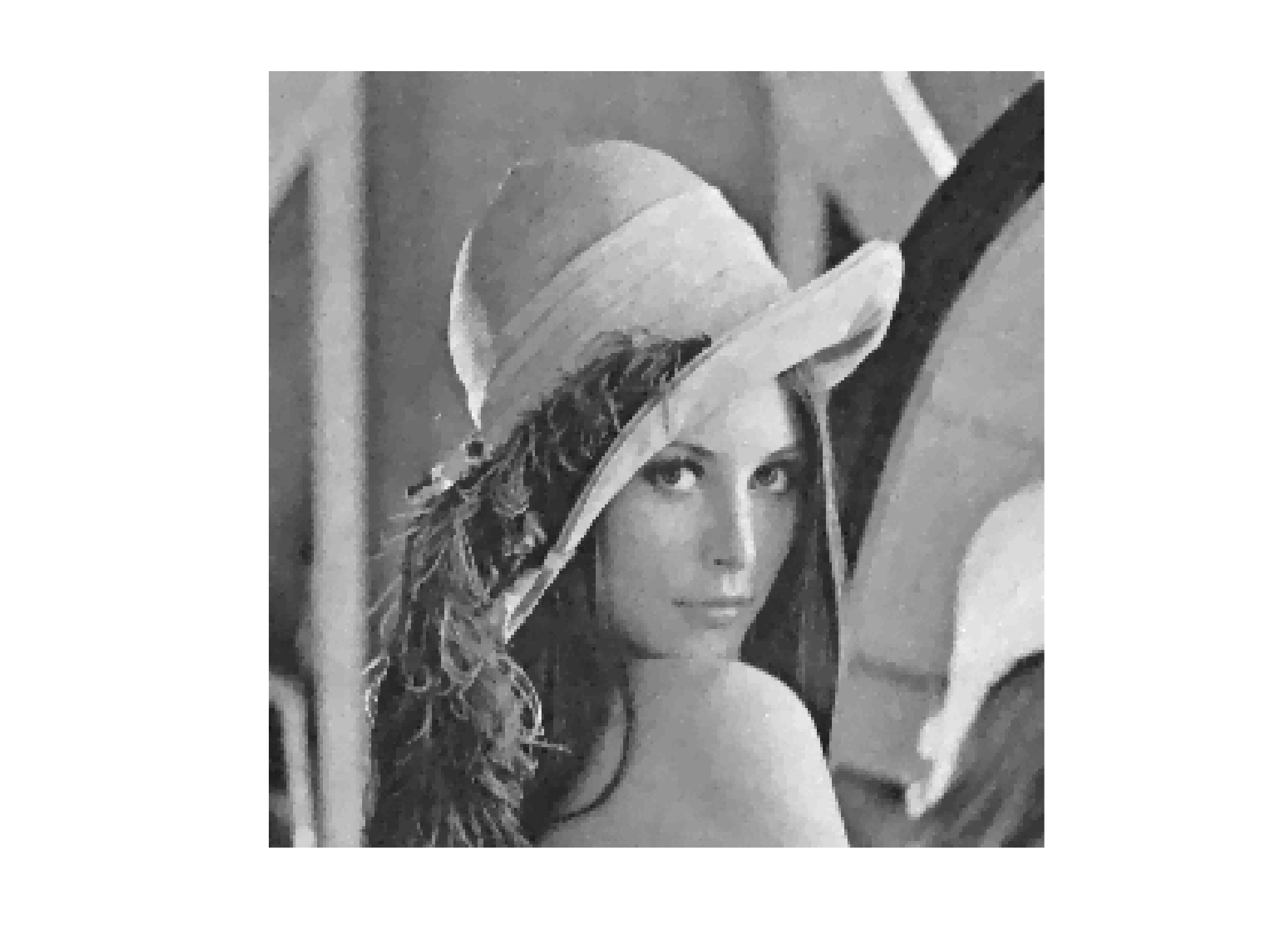}
    \includegraphics[trim = 3.9cm 1.1cm 2cm 1cm, clip, keepaspectratio, height = 3.8cm]{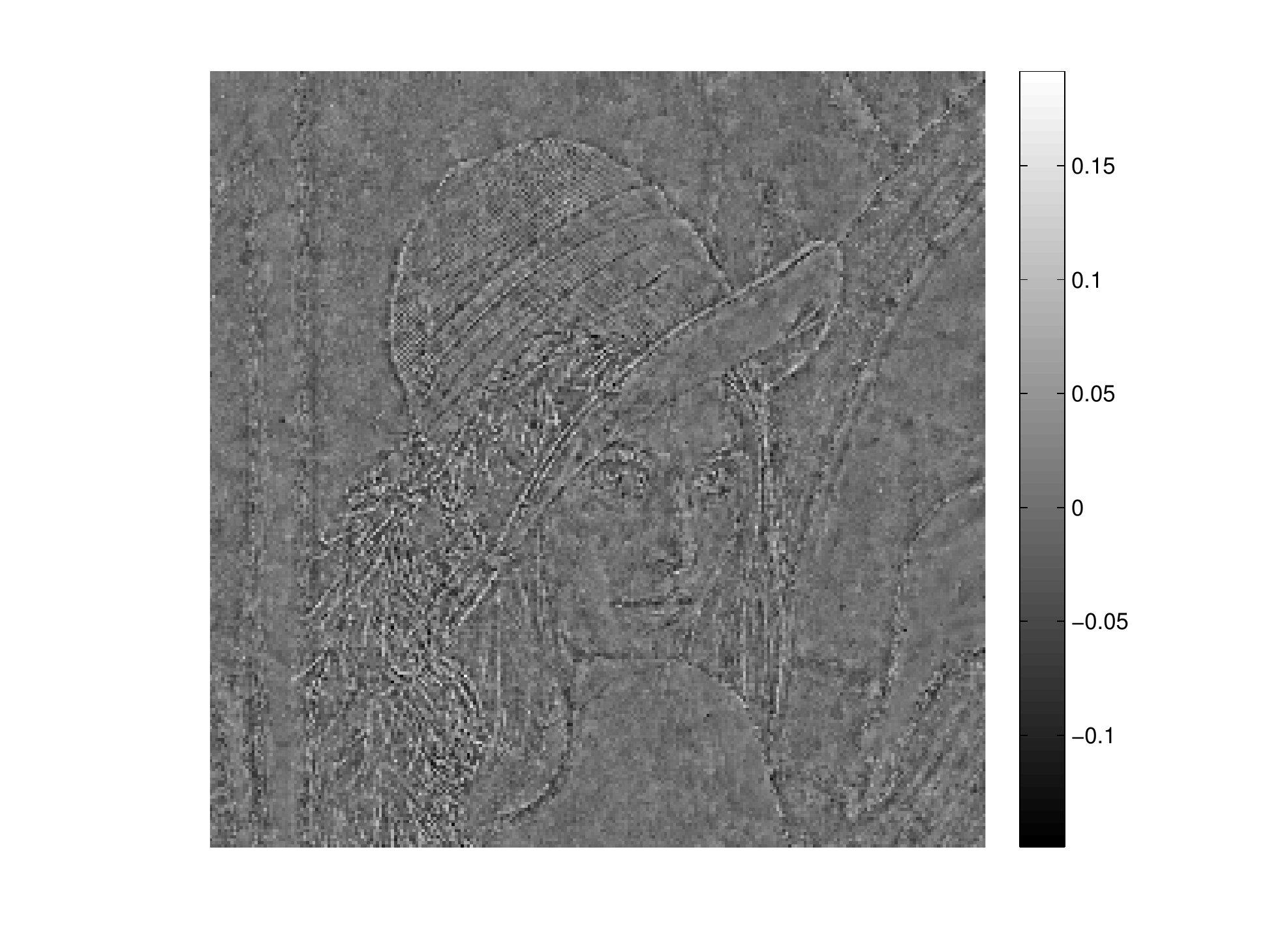}
    \includegraphics[trim = 3.9cm 1.1cm 3cm 1cm, clip, keepaspectratio, height = 3.8cm]{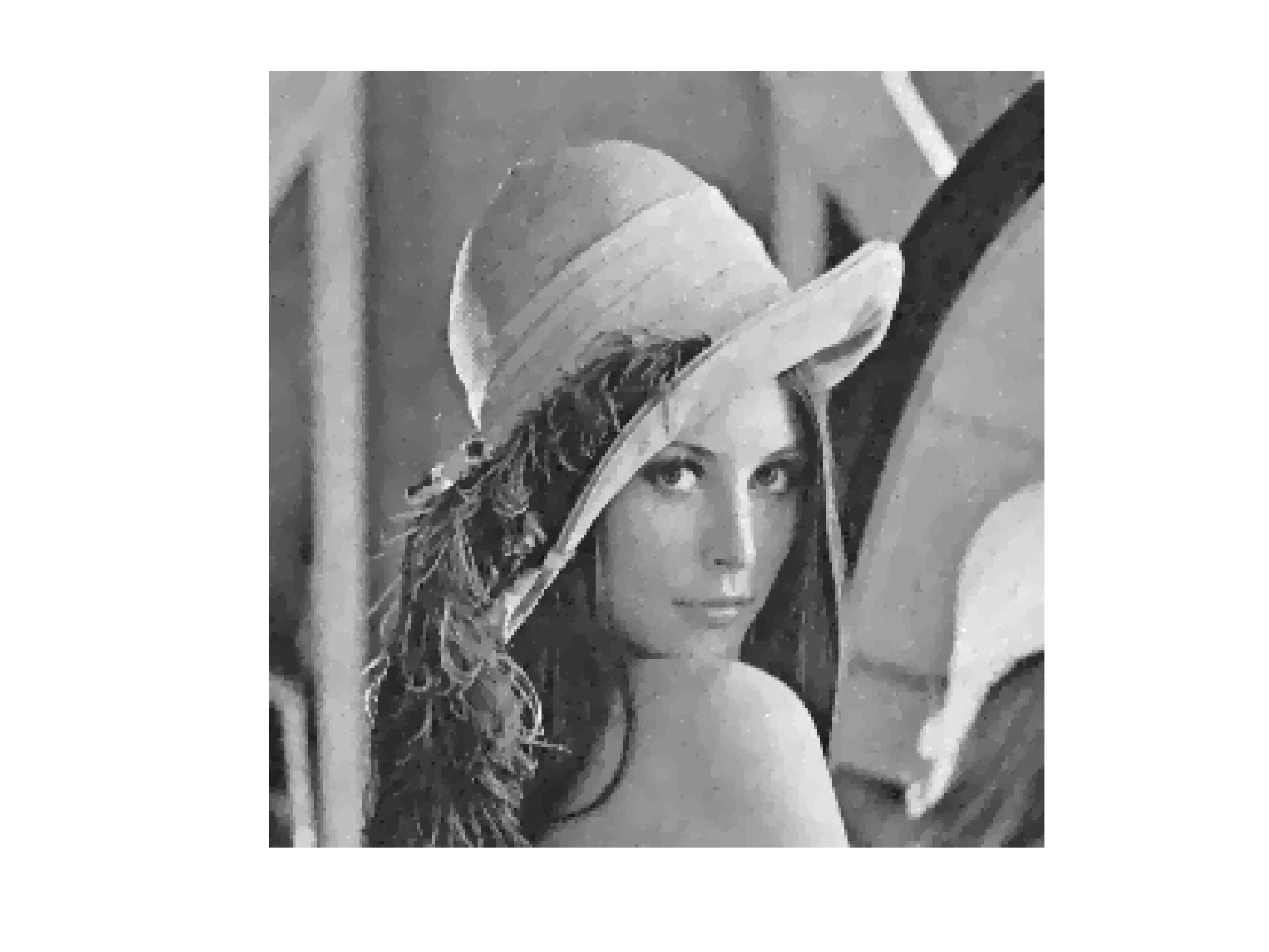}
    \includegraphics[trim = 3.9cm 1.1cm 2cm 1cm, clip, keepaspectratio, height = 3.8cm]{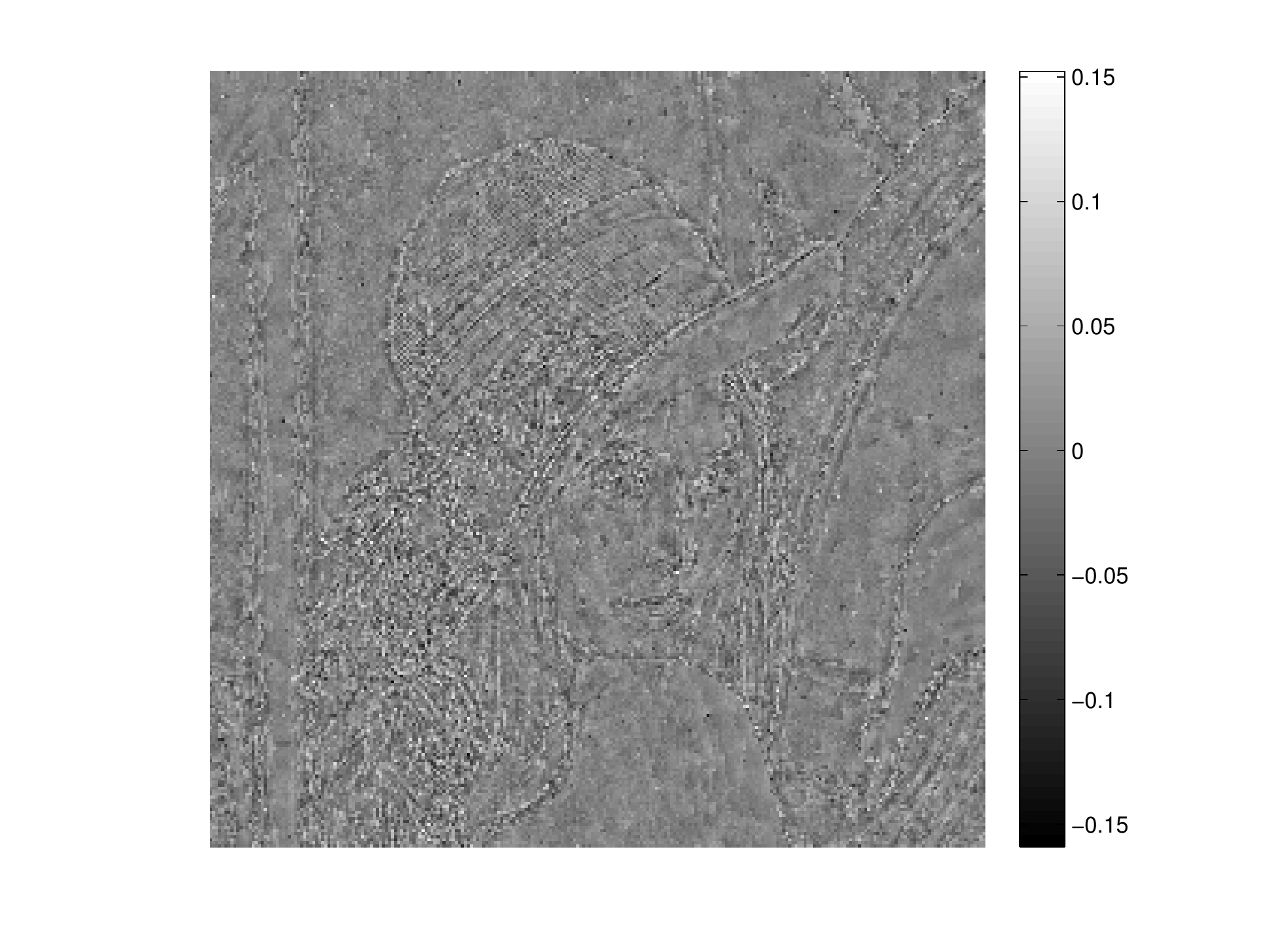}
   
    \includegraphics[trim = 3.9cm 1.1cm 3cm 1cm, clip, keepaspectratio, height = 3.8cm]{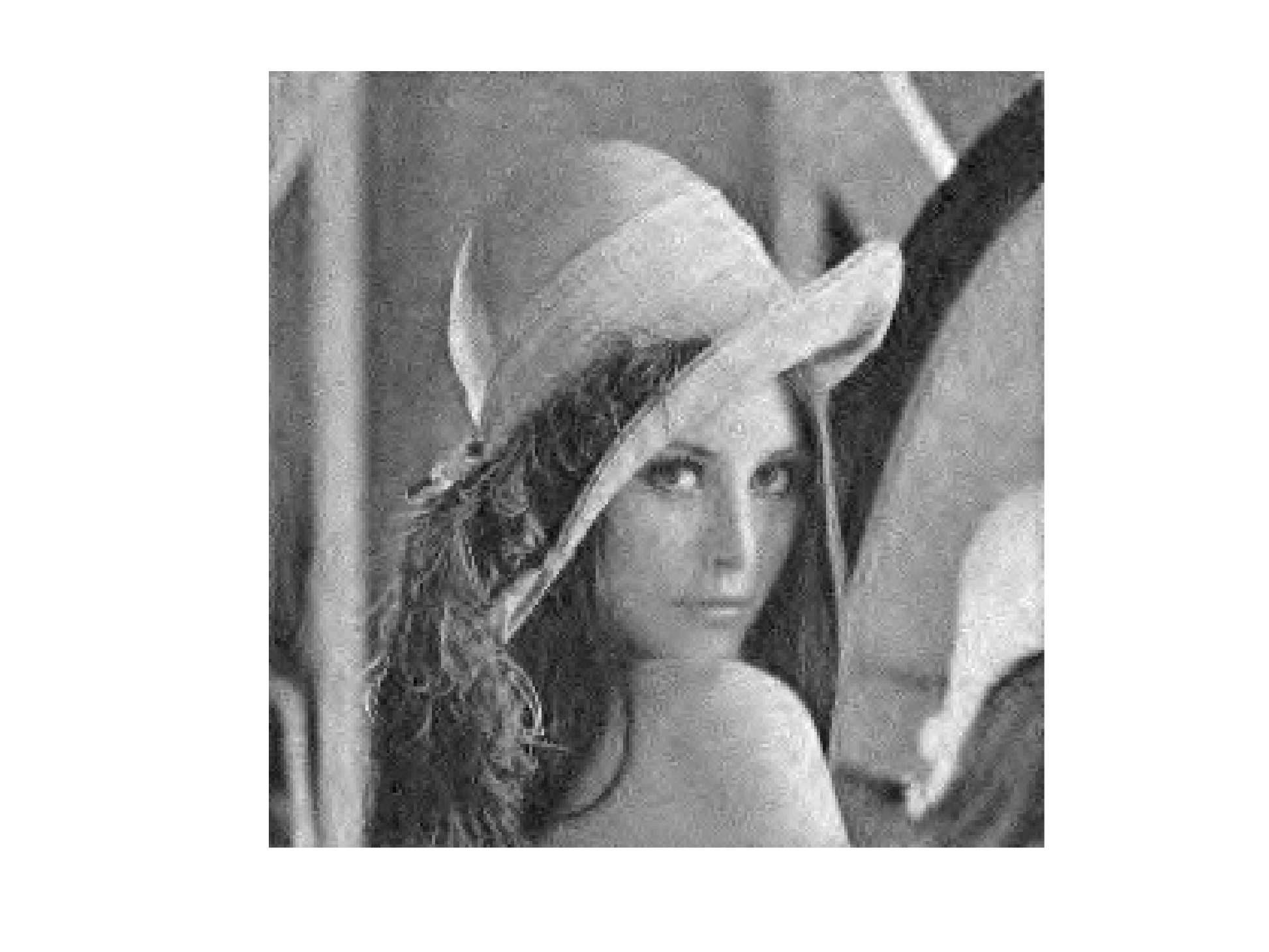}
    \includegraphics[trim = 3.9cm 1.1cm 2cm 1cm, clip, keepaspectratio, height = 3.8cm]{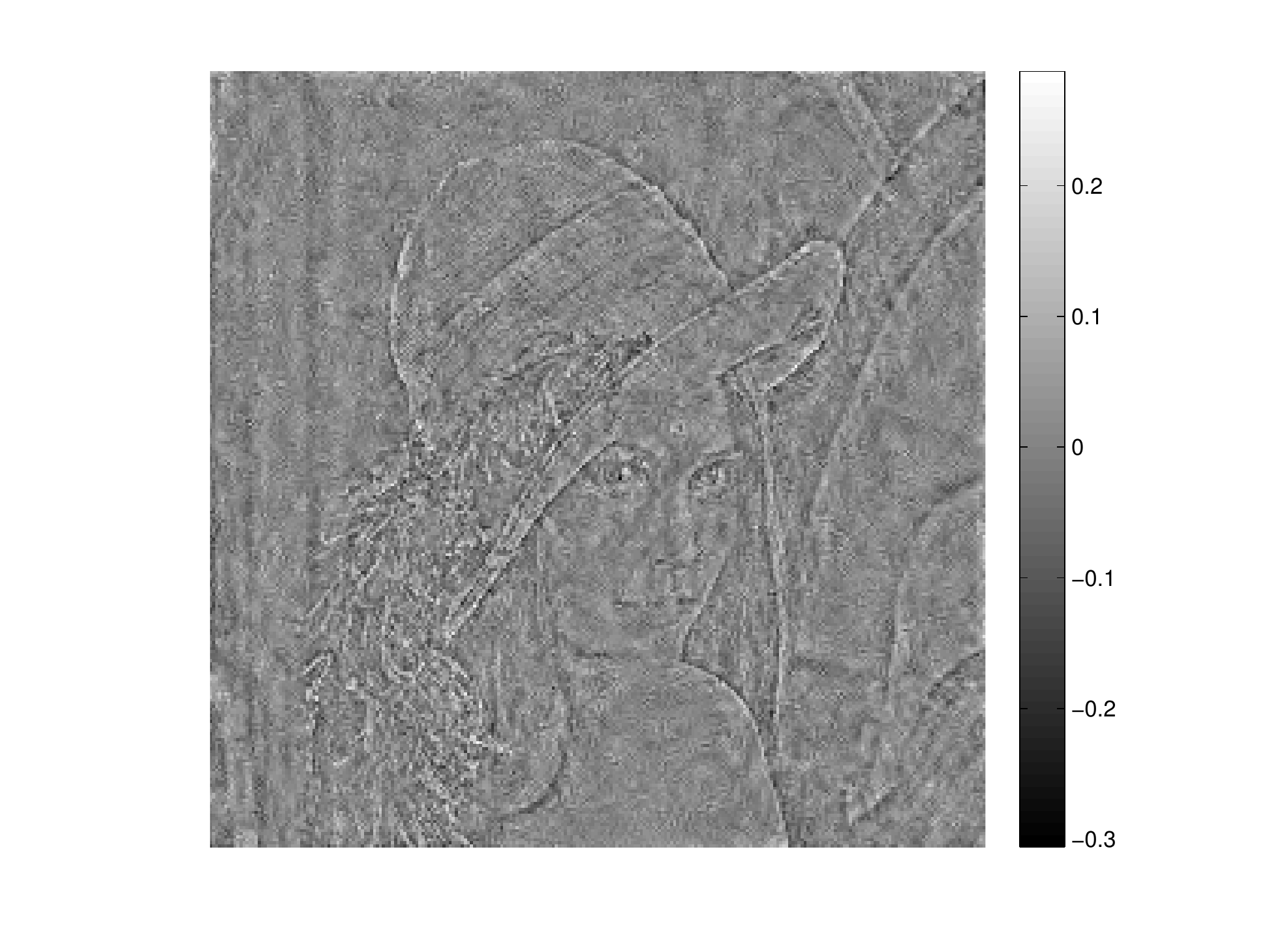}
    \includegraphics[trim = 3.9cm 1.1cm 3cm 1cm, clip, keepaspectratio, height = 3.8cm]{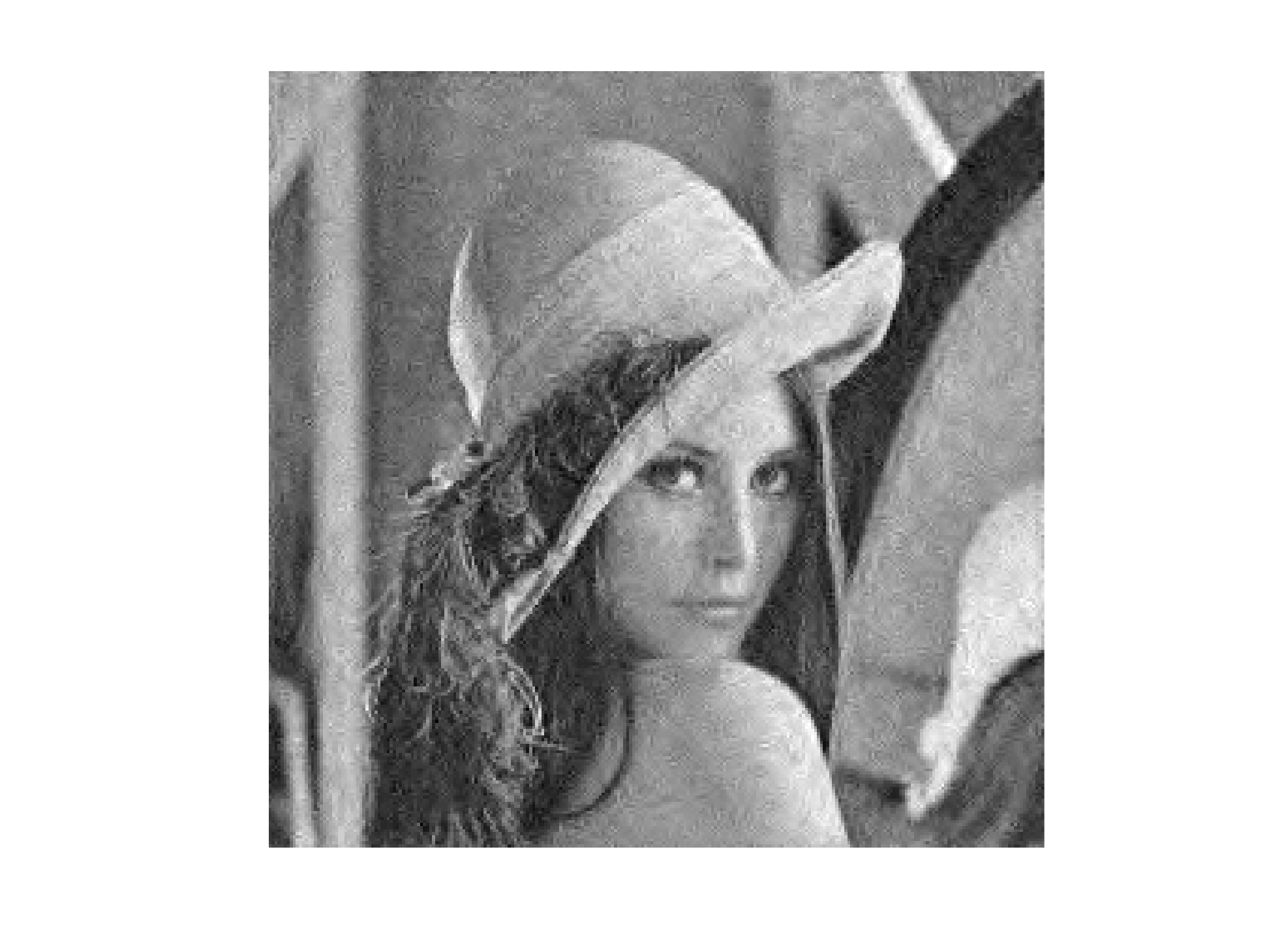}
    \includegraphics[trim = 3.9cm 1.1cm 2cm 1cm, clip, keepaspectratio, height = 3.8cm]{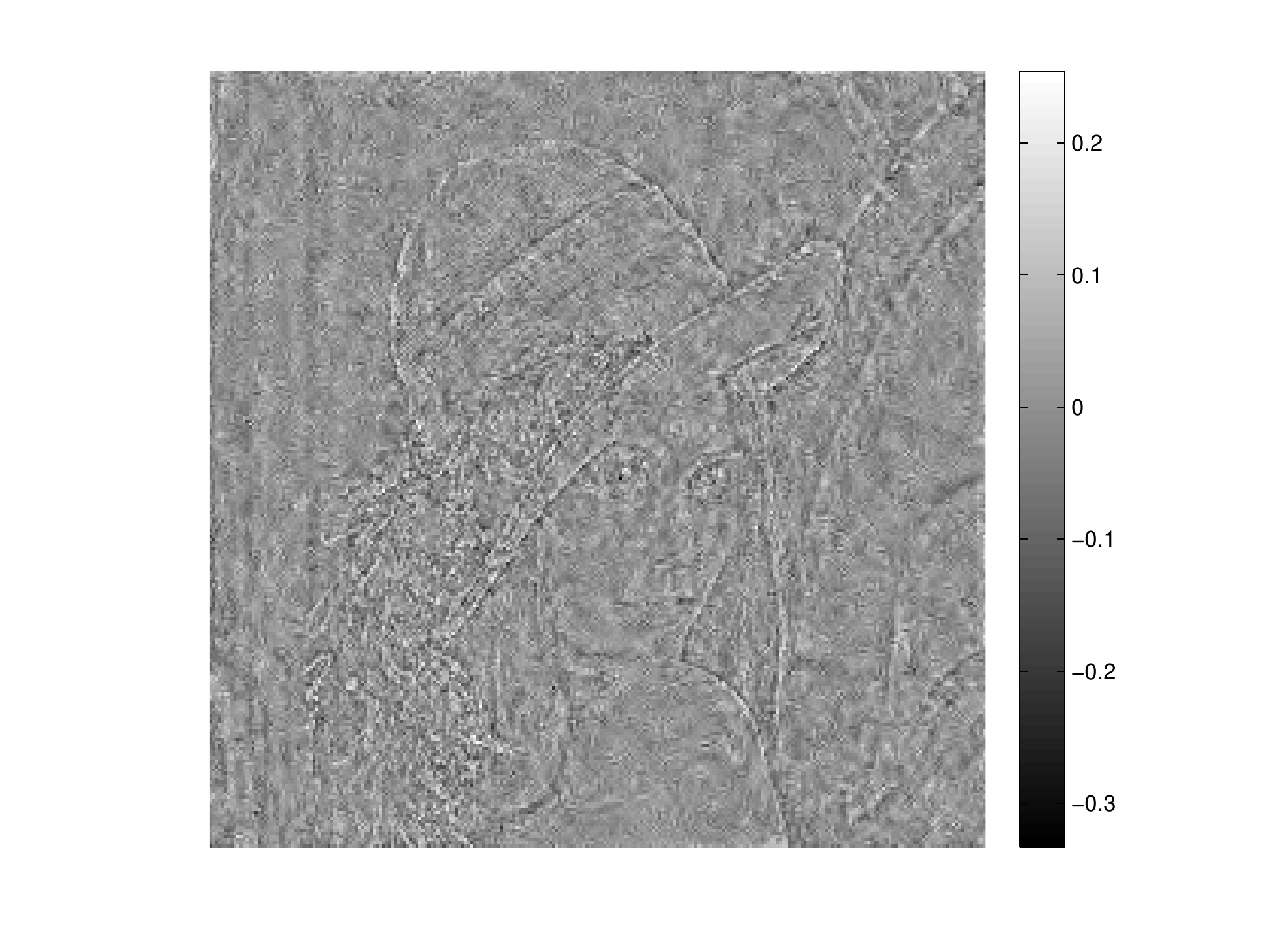}

    \includegraphics[trim = 3.9cm 1.1cm 3cm 1cm, clip, keepaspectratio, height = 3.8cm]{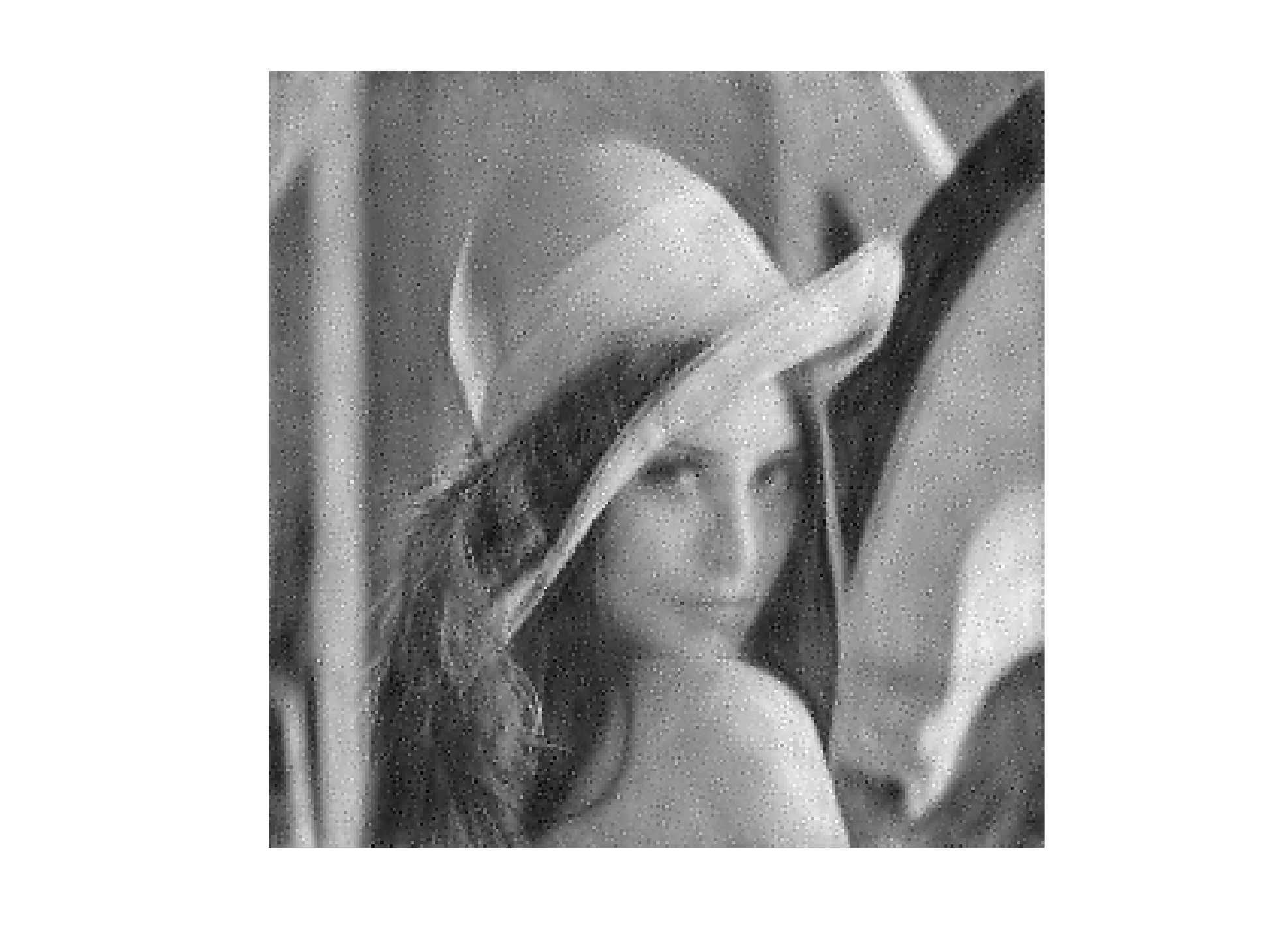}
    \includegraphics[trim = 3.9cm 1.1cm 2cm 1cm, clip, keepaspectratio, height = 3.8cm]{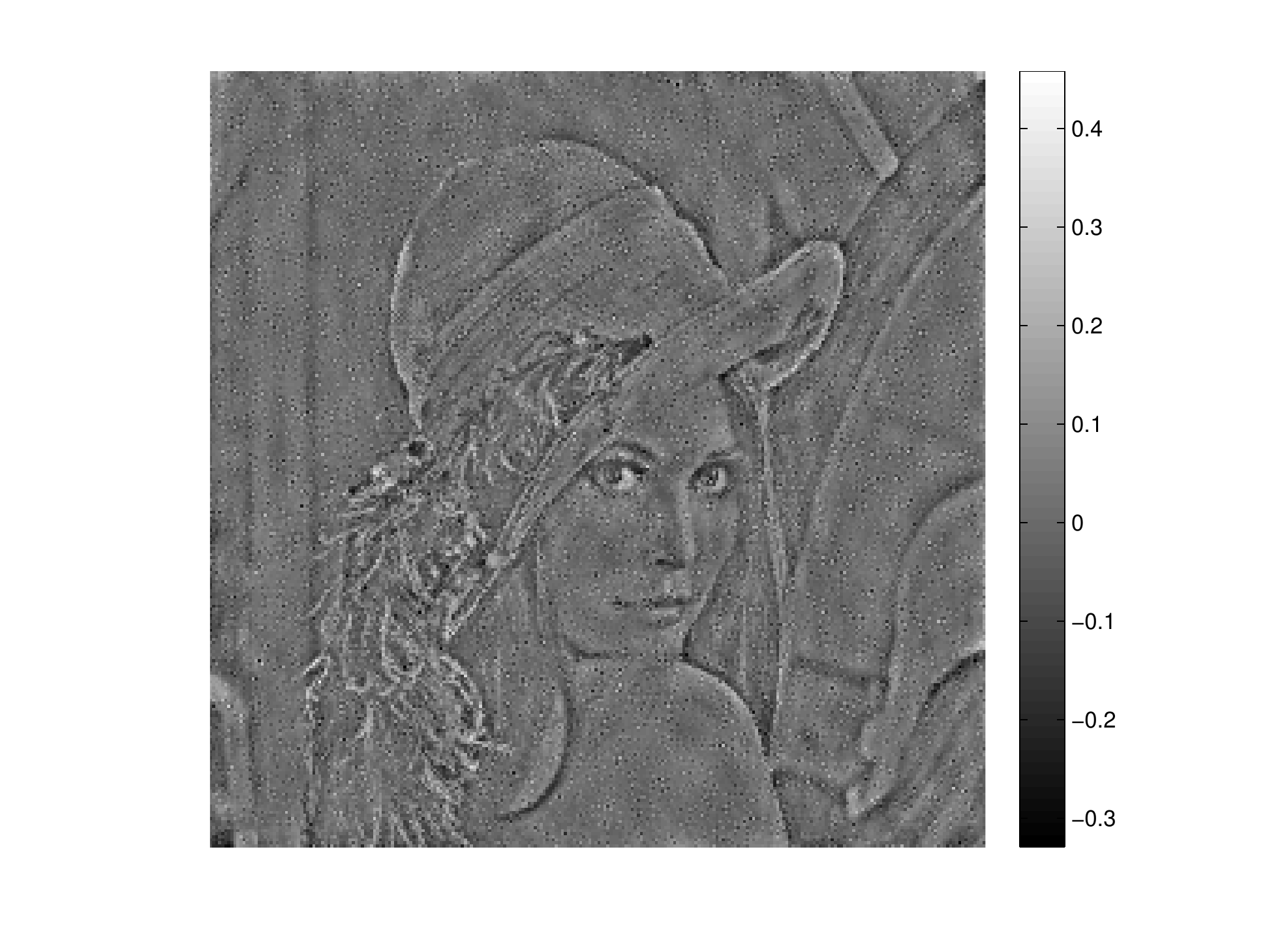}
    \includegraphics[trim = 3.9cm 1.1cm 3cm 1cm, clip, keepaspectratio, height = 3.8cm]{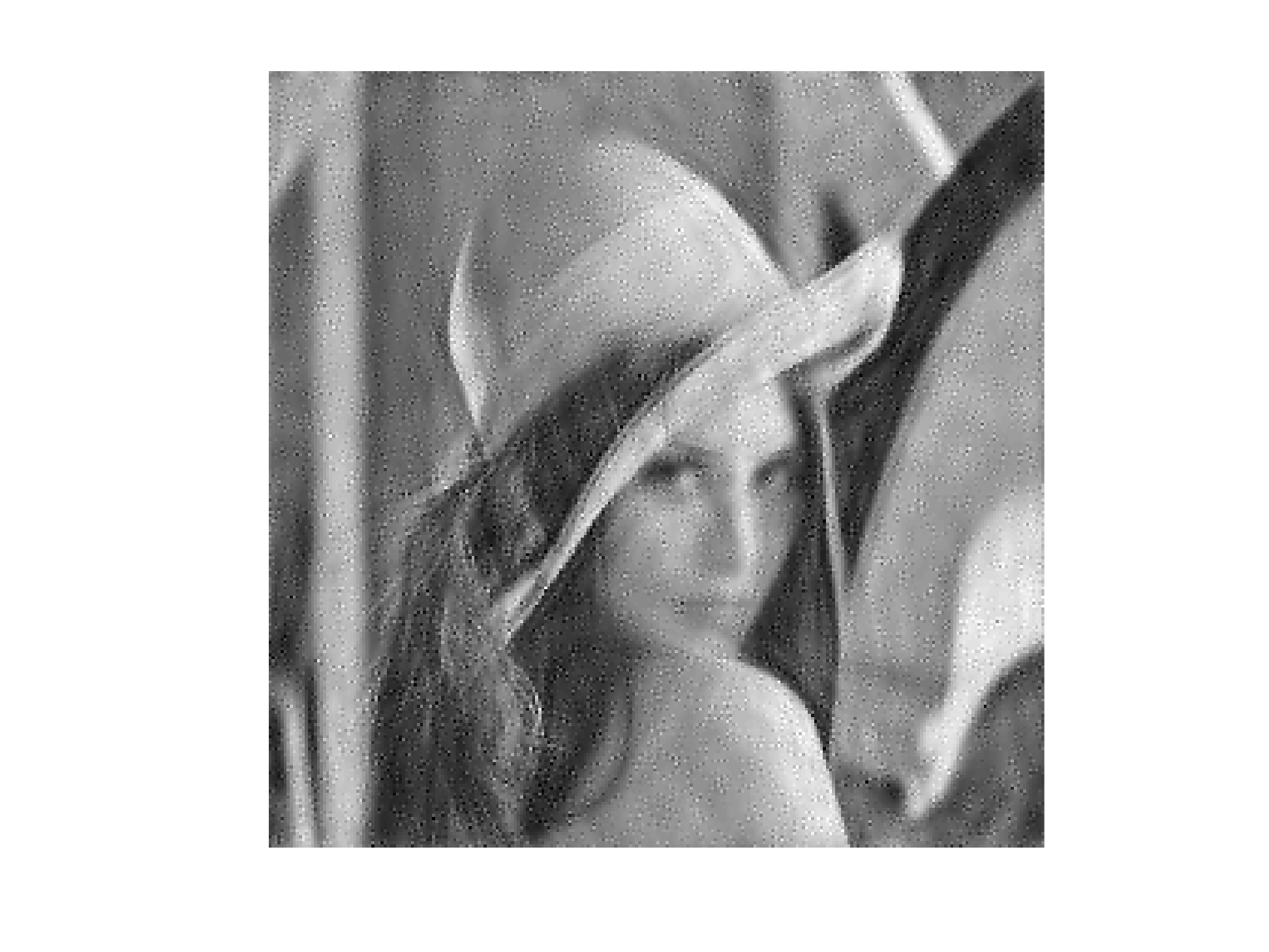}
    \includegraphics[trim = 3.9cm 1.1cm 2cm 1cm, clip, keepaspectratio, height = 3.8cm]{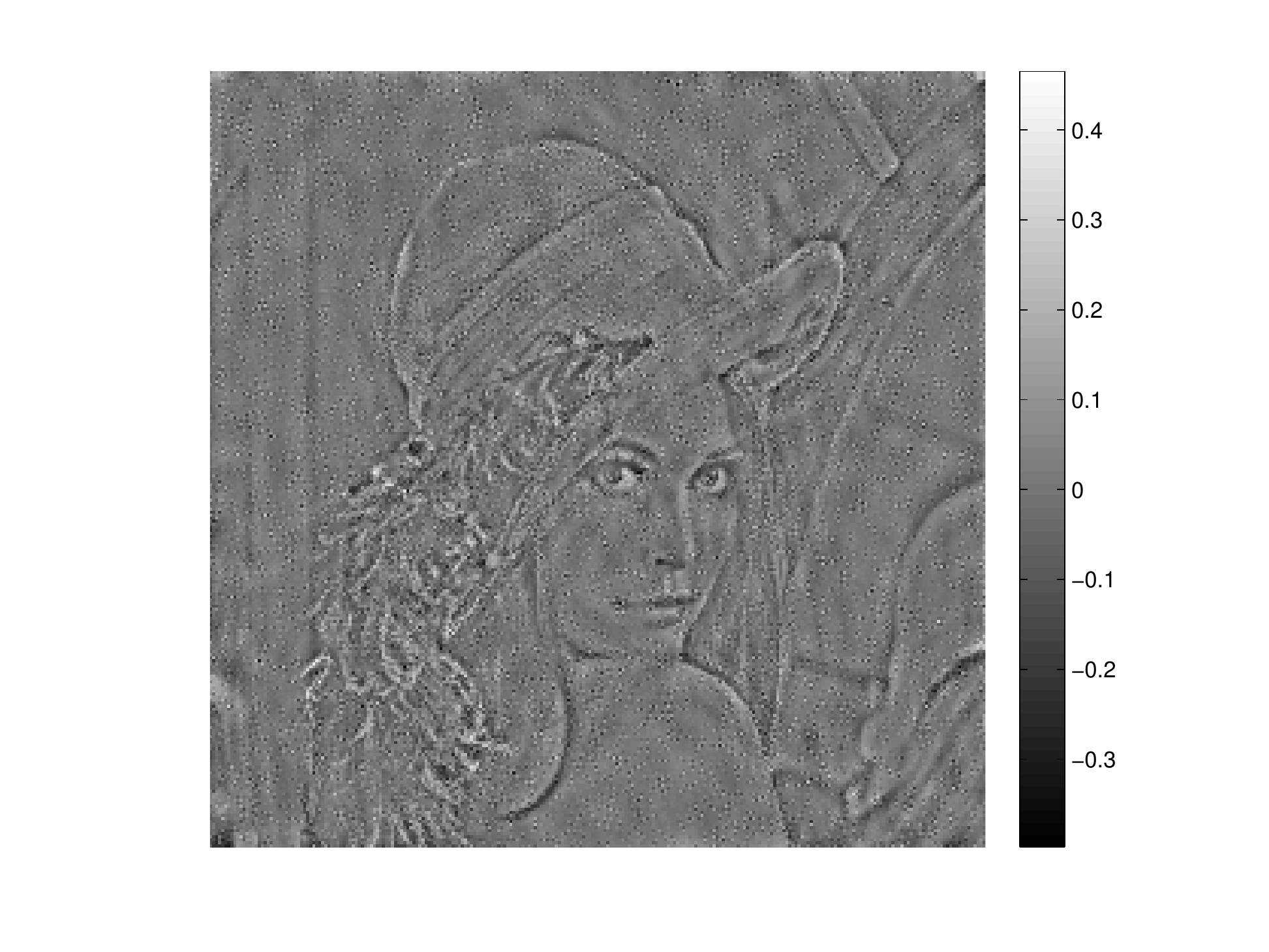}

\caption{Reconstruction example of Lena for spread spectrum acquisition, with $M=0.2N$ and input SNR set to 30~dB. First and third columns show the reconstructed images and the second and fourth columns show the error images. First row: BPSA(24.4~dB) and SARA (27.9~dB). Second row: TV(26.3~dB) and RW-TV (26.6~dB). Third row: BPDb8 (21.4~dB) and RW-BPDb8 (21.2~dB). Fourth row: Curvelet (18.7~dB) and RW-Curvelet (18.3~dB).}
\label{fig:2}
\end{figure*}

In this section, the superiority of SARA to regularization methods based on sparsity in a single frame, as established through simulations in the context of a generic spread spectrum acquisition, is described with a new extensive visual support. Moreover, we bring a novel illustration for a realistic continuous Fourier sampling strategy of particular interest for radio interferometry.

For the first experiment we recover a $256\times256$ version of Lena from compressive measurements. The spread spectrum technique described in \cite{puy12b} is used as measurement operator. We compare SARA to analogous analysis algorithms, and their reweighted versions, changing the sparsity dictionary $\mathsf{\Psi}$ in \eqref{cs6} and \eqref{delta} respectively. Three different dictionaries are considered: the Daubechies 8 wavelet basis, the redundant curvelet frame~\cite{starck10} and the concatenation of the first eight Daubechies bases described above for SARA. The associated algorithms are respectively denoted BPDb8, Curvelet and BPSA for the non reweighted case. The reweighted versions are respectively denoted RW-BPDb8, RW-Curvelet and SARA. Additionally, we also compare to the TV prior, where the TV minimization problem is formulated as a constrained problem like \eqref{cs6}, but replacing the $\ell_1$ norm by the image TV norm. The reweighted version of TV is denoted as RW-TV. Since the image of interest is positive, we impose the additional constraint that $\bar{\bm{x}}\in\mathbb{R}_{+}^N$ for all problems. The reconstruction quality of SARA is evaluated as a function of the undersampling ratio $M/N$, for $M/N$ in the range $[0.1,0.9]$. The input SNR is set to 30 dB. The SNR results comparing SARA against all the other benchmark methods are shown in the right panel of Figure \ref{fig:1}. The results demonstrate that SARA outperforms state-of-the-art methods for all undersampling ratios. RW-TV provides the second best results. BPSA achieves better SNRs than BPDb8, curvelet and their reweighted versions for all undersampling ratios. It also achieves similar SNRs to TV in the range 0.4-0.9. Figure \ref{fig:2} presents a visual assessment for $M=0.2N$, showing both reconstructed and error images. SARA provides an impressive reduction of visual artifacts relative to the other methods in this high undersampling regime. In particular RW-TV exhibits expected cartoon-like artifacts. Other methods  do not yield results of comparable quality, either in SNR or visually, with associated reconstructions full of visual artifacts.

\begin{figure}[t]
    \centering
    \includegraphics[trim = 3.4cm 1.1cm 2.2cm 1.0cm, clip, keepaspectratio, width = 4.3cm]{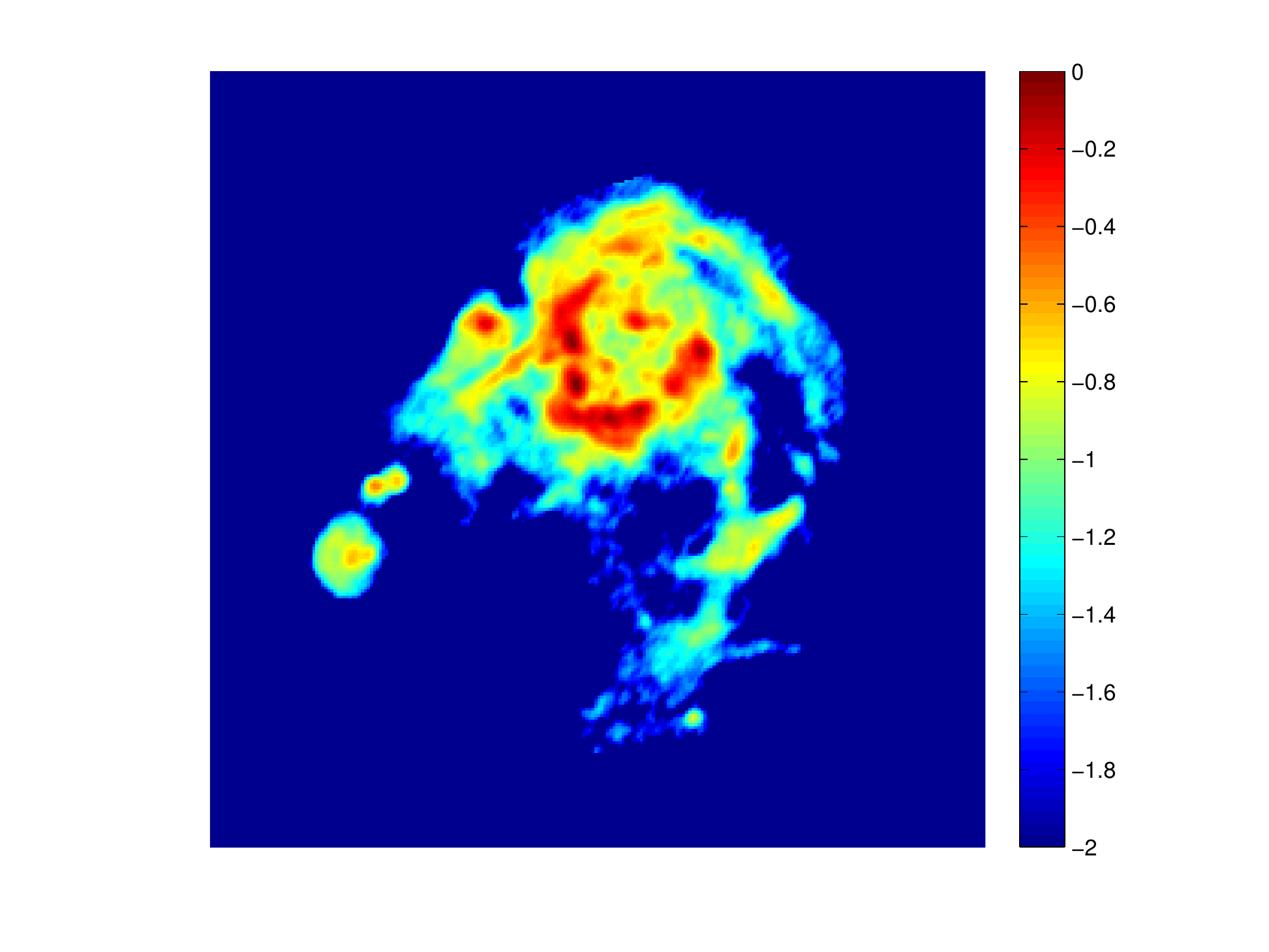}
    \includegraphics[trim = 3.4cm 1.1cm 2.2cm 1.0cm, clip, keepaspectratio, width = 4.3cm]{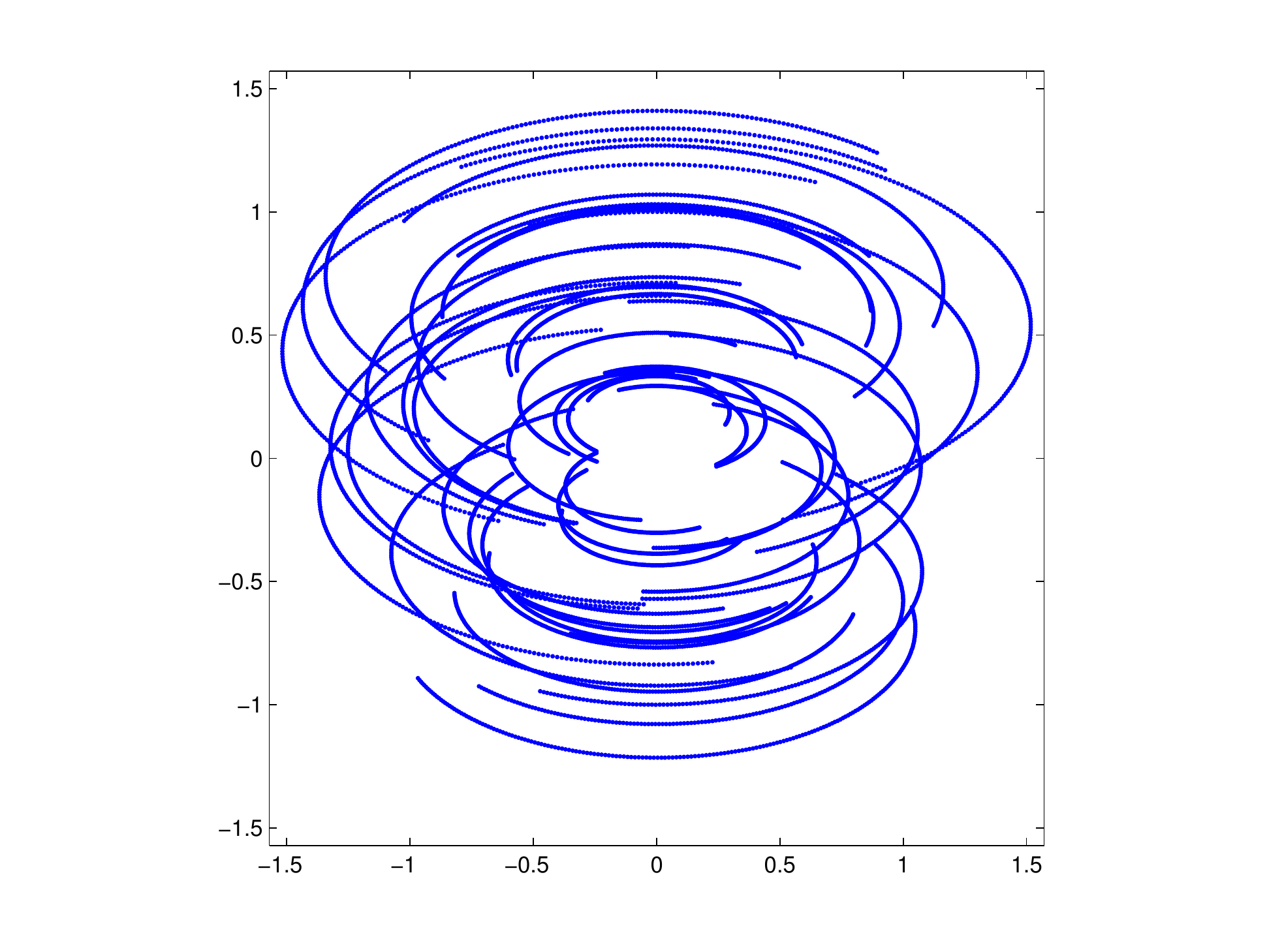}
    \includegraphics[trim = 3.4cm 1.1cm 2.2cm 1.0cm, clip, keepaspectratio, width = 4.3cm]{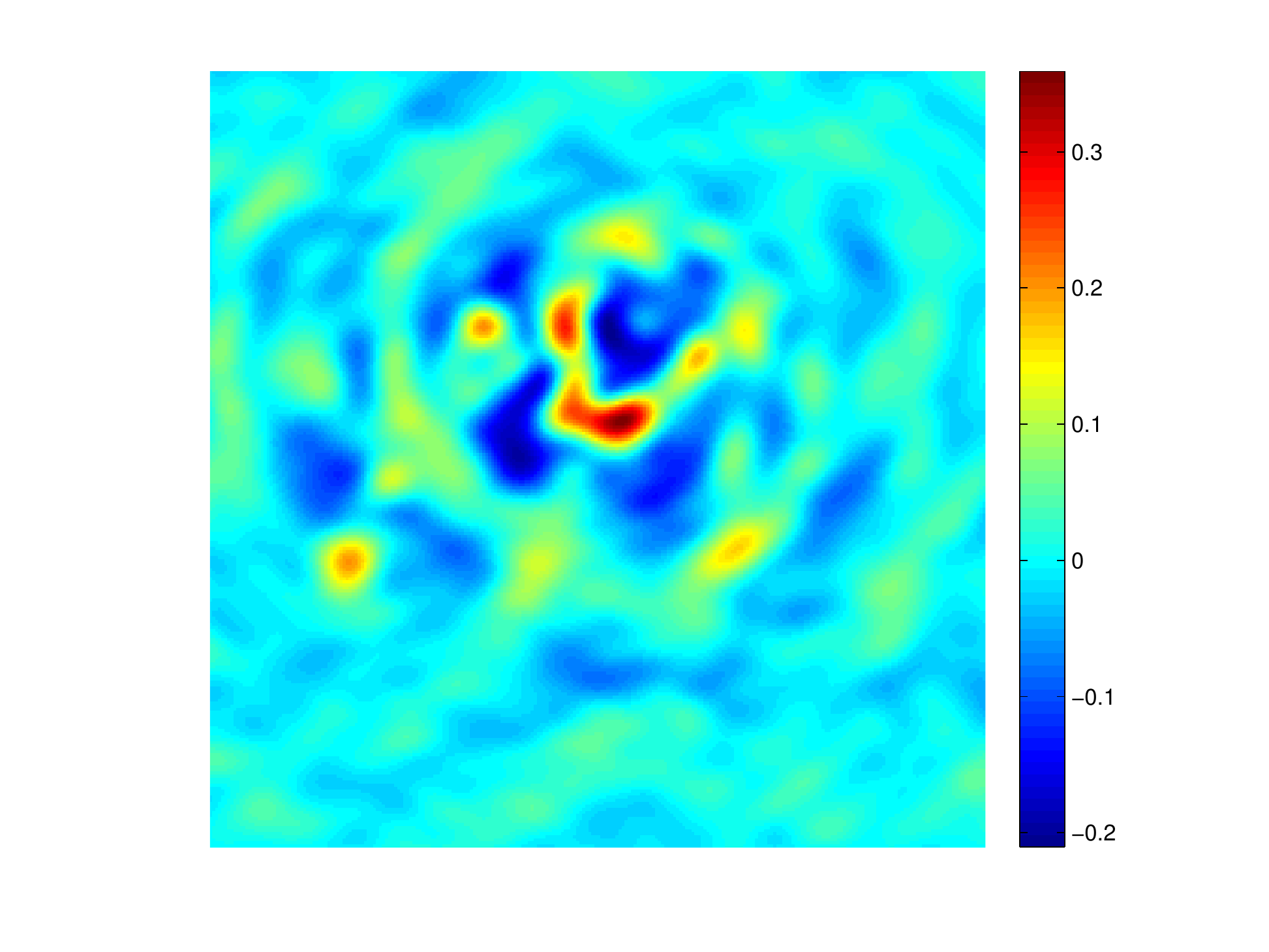}
    \includegraphics[trim = 3.4cm 1.1cm 2.2cm 1.0cm, clip, keepaspectratio, width = 4.3cm]{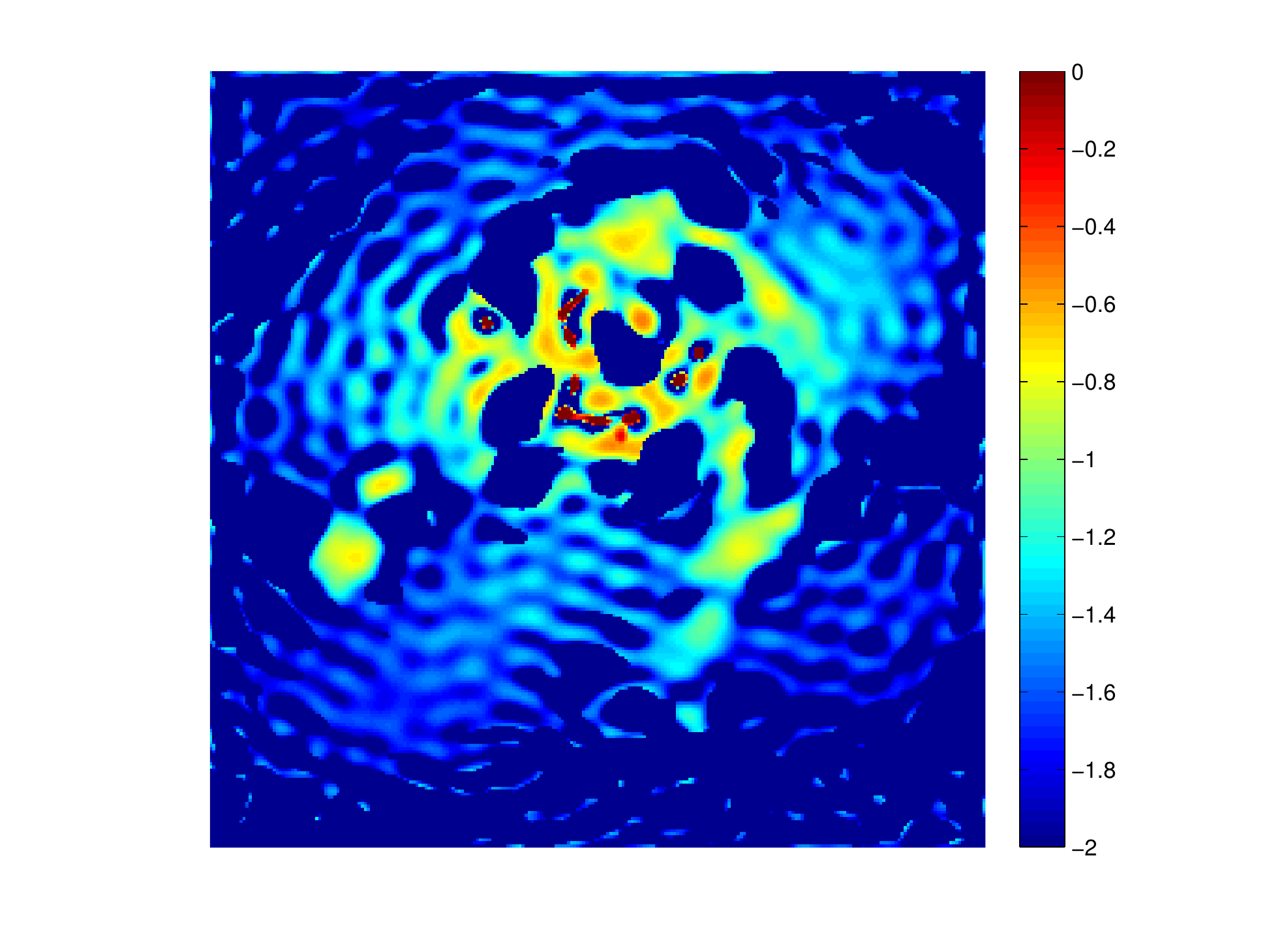}
    \includegraphics[trim = 3.4cm 1.1cm 2.2cm 1.0cm, clip, keepaspectratio, width = 4.3cm]{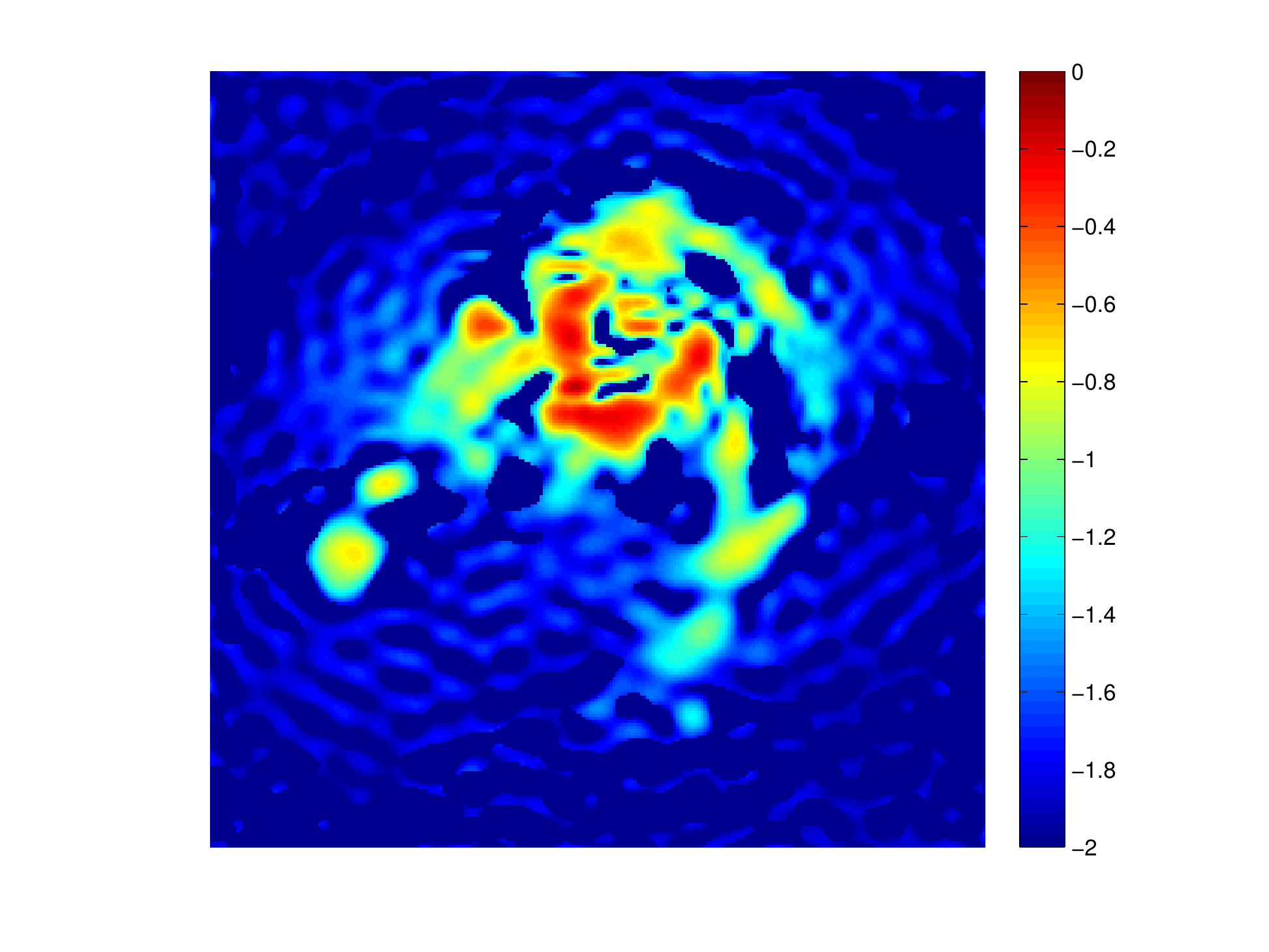}
    \includegraphics[trim = 3.4cm 1.1cm 2.2cm 1.0cm, clip, keepaspectratio, width = 4.3cm]{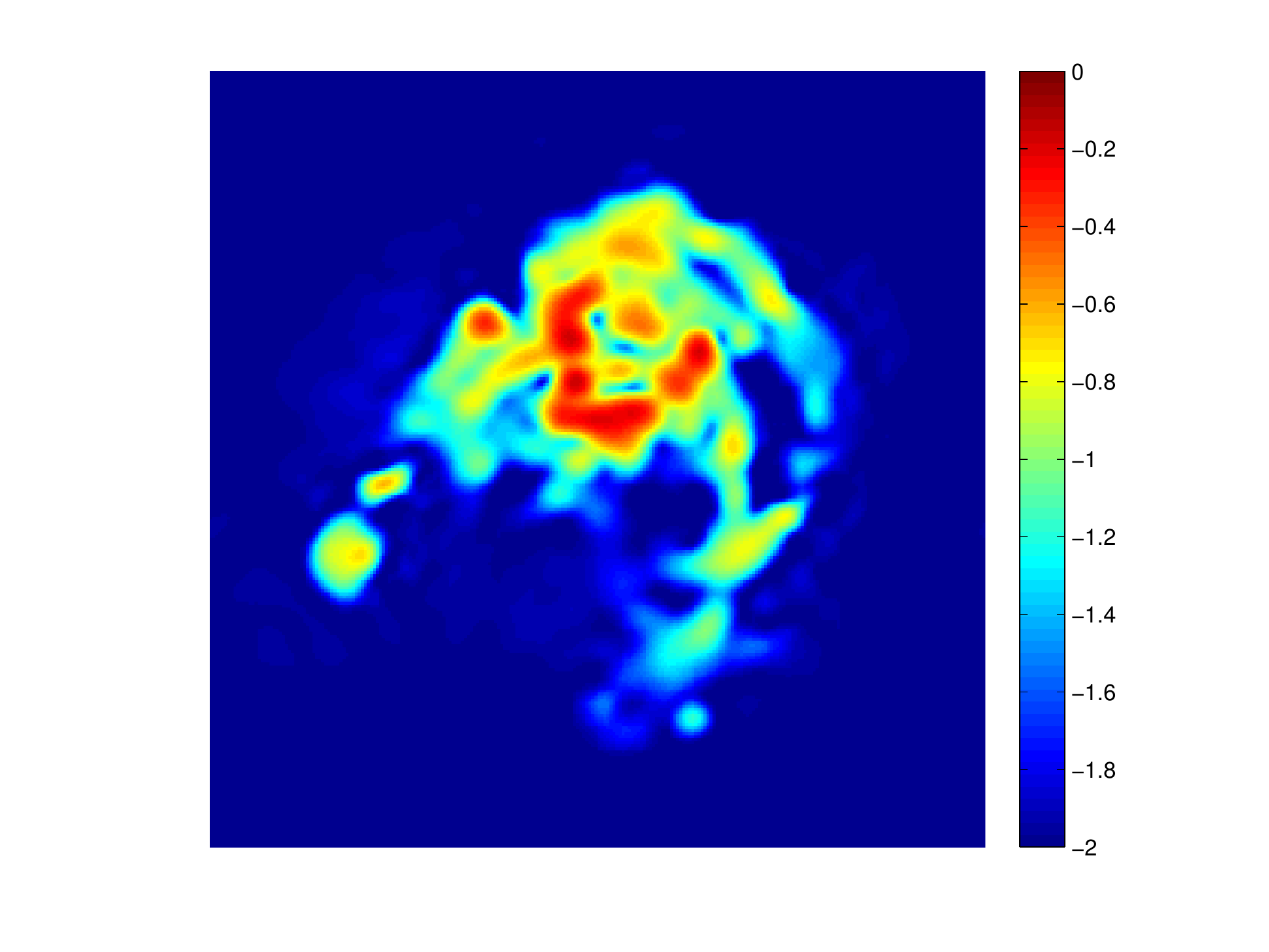}

\caption{Radio Astronomy example. From left to right. Top row: original test image in $\log_{10}$ scale and Fourier sampling pattern. Middle row: corresponding dirty image in linear scale and reconstruction results for BP (3.9~dB) in $\log_{10}$ scale. Bottom row: reconstruction results for BPDb8 (10.3~dB) and SARA (14.1~dB) in $\log_{10}$ scale.}
\label{fig:3}
\end{figure}

The second experiment illustrates the performance of SARA in the context of radio interferometric imaging by recovering a $256\times256$ version of the well known M31 galaxy from simulated continuous Fourier samples associated with a realistic radio telescope sampling pattern (superposition of arcs of ellipses). The number of measurements is $M=9413$, affected by 30 dB of input noise. The dictionary for SARA is the concatenation of the first eight Daubechies bases \emph{and} the Dirac basis. The Dirac basis is added given the sparsity in image space due to the large field of view.  For comparison, we use two different methods: BP, constrained $\ell_1$-minimization in the Dirac basis (used as benchmark in the field), and BPDb8, constrained analysis-based $\ell_1$-minimization in the Db8 basis. Figure~\ref{fig:3} shows the original test image, the sampling pattern and the corresponding dirty image, i.e.~the inverse Fourier transform of the measurements, with non-measured points set to zero. The reconstructed images for BP, BPDb8 and SARA are also reported. Once more, SARA provides not only a drastic SNR increase but also a significant reduction of visual artifacts relative to the other methods. 

\section{Conclusion and Discussion}

In this paper we have reviewed recent advances in the average sparsity model and the associated algorithm SARA. Extended simulations demonstrating the superiority of SARA for compressive imaging reconstruction were described. Novel results on the application of SARA to a realistic radio interferometric imaging scenario were also described.

Future work will concentrate on finding a theoretical framework for the average sparsity model. In \cite{carrillo13} we have put average sparsity in the context of theory developed in \cite{candes10}. However, specialized results for the particular case of concatenation of frames (or orthogonal bases) are needed. The co-sparsity analysis model \cite{nam13} proposes a general framework for general analysis operators. Similar properties to the D-RIP coined $\Omega$-RIP are introduced in \cite{giryes13} to analyze greedy algorithms in the context of the co-sparsity analysis model. It would be interesting to explore the connections between average sparsity and the co-sparsity model to have an estimate on the number of measurements needed for reconstruction compared to single frame representations.

The proposed approach relies on the observation that natural images exhibit strong average sparsity, i.e.~the signals of interest have so-called simultaneous structured models. Recently, it was shown in \cite{oymak13} that combinations of convex relaxations of the individual structured models do not yield better results than an algorithm that exploits only one of the structured models, while \emph{non-convex} approaches that approximate the simultaneous model can exploit the multiple structured models. Those results suggest that the \emph{re-weighting} approach in SARA to approximate the $\ell_0$ norm is fundamental to exploit average sparsity, as observed in the simulation results (see the gap between SARA and BPSA in Fig.~\ref{fig:1} and Fig.~\ref{fig:2}).

\section*{Acknowledgment}
REC is supported by the Swiss National Science Foundation (SNSF) under grant 200021-130359. JDM is supported by a Newton International Fellowship from the Royal Society and the British Academy. YW is supported by the Center for Biomedical Imaging (CIBM) of the Geneva and Lausanne Universities and EPFL.

\bibliographystyle{IEEEtran}
\bibliography{abrev,sara}

\end{document}